\documentclass[aps,prx,groupedaddress,twocolumn,showpacs,longbibliography,10pt]{revtex4-2}

\usepackage{amsmath}
\usepackage{amsthm,amssymb,bbold}
\usepackage{bm}
\usepackage{braket}
\usepackage{graphicx}
\usepackage{color}
\usepackage{mathtools,tikz}
\usepackage{natbib}
\usepackage{hyperref}
\usepackage{lipsum}

\hypersetup{
       colorlinks=true,
}

\newcommand{\CO}{{\cal O}}
\usepackage{etoolbox}

\makeatletter
\newcommand*{\rom}[1]{\expandafter\@slowromancap\romannumeral #1@}
\makeatother

\date{\today}

\begin{document}

\title{ Certain BCS wavefunctions are quantum many-body scars }

\author{Kiryl Pakrouski$^{1}$, Zimo Sun,$^{2}$}
\affiliation{$^{1}$Institute for Theoretical Physics, ETH Zurich, 8093 Zurich, Switzerland}
\affiliation{$^{2}$Department of Physics, Princeton University, Princeton, NJ 08544, USA}

\begin{abstract}
We construct many-body scar states in multi-flavour fermionic lattice models that possess strong magnetic or superconducting correlations of a given type specified by a unitary matrix $A$. One of the states maximizes the one-point correlations over the full Hilbert space and has the form of the BCS wavefunction. It may always be made the ground state by adding the correlations as a "pairing potential" to any Hamiltonian supporting group-invariant scars. In our single-flavour, spin-full fermions example we consider a superconducting $A$. The BCS scar ground state is a linear combination of the well-known $\eta$-pairing states. In the multi-orbital fermions example the BCS-like ground state maximizes unconventional magnetic correlations. The broad class of eligible Hamiltonians includes many conventional condensed matter interactions. The part of the Hamiltonian that governs the exact dynamics of the scar subspace coincides with the BCS mean-field Hamiltonian. We therefore show that its eigenstates are many-body scars that are decoupled from the rest of the Hilbert space and thereby protected from thermalization. Our results point out a connection between the fields of superconductivity and weak ergodicity breaking (many-body scars) and will hopefully encourage further investigations. They also provide the first feasible protocol to initialize a fermionic system to a scar state in (a quantum simulator) experiment.

\end{abstract}

\maketitle

Superconductivity \cite{TinkhamBookIntroSC} is one of the best-understood quantum phenomena which still conceals some of the most intriguing secrets such as the mechanisms underlying unconventional superconductivity. 
As a macroscopic quantum phenomenon, superconductivity reflects the collective behaviour of many degrees of freedom. The emergence of statistical physics from individual quantum eigenstates can be analyzed in terms of eigenstate thermalization hypothesis (ETH) \cite{deutsch1991quantum,srednicki1994chaos,rigol2008thermalization} that conjectures that a macroscopic thermal average of an observable should agree with measurements made in individual eigenstates at the corresponding energy.

Many-body scars (MBS) are eigenstates of strongly-interacting Hamiltonians that elude this description \cite{RydbergExperimentRevivals,Serbyn:2020wys,Moudgalya:2021xlu,Papic2022,Chandran:2022jtd} by weakly breaking ergodicity. They are found in a variety of systems \cite{PhysRevA.86.041601,Shiraishi2017ScarsConstruction,Turner_2018,Moudgalya:2018,AbaninScarsSU2Dynamics,Khemani:2019vor,Sala_2020,Prem:2018,Schecter:2019oej,BucaNature2019,SciPostPhys.3.6.043,IadecolaHubbardAlmostNUPRL2019,Shibata:2020yek,michailidis2020stabilizing,2020MarkMotrEtaPairHub,PRLPapicClockModels,VedikaScarsVsIntegr,Pal2020ScarsFromFrustration,mark2020unified,iadecola2020quantum,moudgalya2020etapairing,PhysRevB.101.220305,pakrouski2020GroupInvariantScars,PhysRevLett.126.120604,PhysRevResearch.2.043305,Nielsen2020ChiralScars,Hsieh2020PXP2D,Regnault2020MPStoFindSc,FloquetPXPScars,PapicWeaklyBrokenAlgebra,kuno2021multiple,banerjee2020quantum,Pakrouski:2021jon,chaoticDickeAllScarred2021,maskara2021DrivenScars,langlett2021rainbow,2021arXiv210807817R,2021arXiv211011448T,Schindler:2021lma,Dodelson:2022eiz,Liska:2022vrd,scarsInSchwingerModel2023,TruncatedSchwingerModel2023,paper3Case1Majorana,budde2024quantummanybodyscarsarbitrary,NielsenTowerInLocalized,shen2024enhancedmanybodyquantumscars,osborne2024quantummanybodyscarring21d} and may be understood within several frameworks \cite{Shiraishi2017ScarsConstruction,pakrouski2020GroupInvariantScars,Moudgalya:2022nll}. We will think about MBS as a subspace that dynamically decouples from the rest of the Hilbert space and is governed by often simple and integrable Hamiltonian $H_0$ while the Hamiltonian of the full system $H=H_0+OT$ is strongly interacting and has almost no symmetries \cite{pakrouski2020GroupInvariantScars}. For certain simple $H_0$ the time evolution within the scar subspace follows "closed orbits" which manifests as "revivals" in experiments \cite{RydbergExperimentRevivals}.

One way to discuss superconductivity is using the concept of the off-diagonal long-range order (ODLRO), the presence of non-decaying with distance two-point correlations between pair annihilation ($\CO_i$) and creation ($\CO^\dagger_j$) operators at distant sites $i$ and $j$ \cite{etaPairingAsMechanismOfSC}. These correlations can also be used as an indicator of superconductivity induced by an external drive in pump-probe experiments \cite{Kaneko2019pumpProbeEta,Werner2020RevivalsOfZetaStatesMultiband,pwaveScarsInSpinlessFermMazza2022}. The concept of ODLRO was introduced by Yang who also constructed the eta-pairing states \cite{etaPairingYang89} (see \cite{Nakagawa:2022jsg,scFlatBand1,scFlatBand2} for generalizations and relation to superconductivity), eigenstates of the Hubbard model that possess ODLRO \cite{etaPairingYang89,yang1990so}.

Interestingly, these exact states have recently been found to be many-body scars \cite{SciPostPhys.3.6.043,2020MarkMotrEtaPairHub} relevant to several important concepts such as spectrum-generating algebras \cite{moudgalya2020etapairing,mark2020unified} which in turn are relevant in a broader context of open quantum systems \cite{BucaNature2019}. Further, MBS with unconventional pairing have been demonstrated \cite{imai2024quantummanybodyscarsunconventional} albeit in a somewhat artificial fine-tuned model with multi-body interactions.
Could these coincidences hint at a possible relation between the phenomena of many-body scars and superconductivity?

One could at first see a contradiction in the fact that the analysis of superconductivity usually starts from the ground state whereas the MBS are typically scattered around the spectrum.

In this work we construct MBS with arbitrary (as opposed to singlet in eta pairing states) pairing for multi-orbital spin-full fermions. In one basis they look like a generalization of the eta-pairing states. In an alternative basis, selected by the "pairing potential" $\delta H_0=-(\CO^\dagger+\CO)$ Hamiltonian, the lowest energy state is the BCS wavefunction that maximizes the order parameter $\braket{\CO^\dagger_i}$ and that can thus always be made the ground state by adding strong enough $\delta H_0$. In either basis the subspace contains the states with highest ODLRO $\braket{\CO^\dagger_i\CO_j}$ in the Hilbert space.

We discuss two types ("superconducting" and "magnetic") of rather general bilinear fermionic excitation operators $\CO^\dagger$ and a broad class of compatible with MBS fermionic lattice Hamiltonians.
The general scheme is illustrated with two examples. 
In the single-orbital case we consider two superconducting and one magnetic $\CO^\dagger$ in a Hubbard model and establish that the BCS scar ground state and its excitations span the same scar subspace as the well-known eta-pairing states. In the two-orbital case we consider an "inter-orbital", magnetic $\CO^\dagger$ and illustrate the presence of the $(\CO^\dagger)^2$ contribution in the BCS-like ground state for a deformed Hubbard Hamiltonian.

$\delta H_0$ can be thought of as a BCS mean-field Hamiltonian and its solution is known to be related to the coherent states \cite{CohStatesRadcliffe1971,CohStatesOutOfEtaPairingStates1998}. We demonstrate that the corresponding wavefunctions are group-invariant and the mean-field solution is thus endowed with extra stability due to the fact that it is a MBS. Our findings strengthen the case that many-body scars may be related to superconductivity and provide a lead to further exploration in this direction.

\emph{General case}---For the excitations of type \rom{1} we consider the Hilbert space of complex fermions $\{c_{i\alpha}^\dagger,  c_{j\beta} \}=\delta_{\alpha\beta}\delta_{ij}$, where $1\le i, j\le N$ are site indexes on an arbitrary lattice of arbitrary dimension and $1\le \alpha, \beta\le 2K$ are the flavour indexes.

Let $A$ be a $2K\times 2K$ antisymmetric and unitary matrix and $n_j$ the particle number operator at site $j$. 
The bilinear of type \rom{1}
\begin{align}
\label{eq:type1O}
\CO_j = \frac{1}{2} \sum_{\alpha, \beta=1}^{2K}c^\alpha_{j} A_{\alpha\beta} c^\beta_{j}; \quad
\left[\CO_j , \CO_j ^\dagger \right] = K-n_j,
\end{align}
creates excitations over the state without any fermions $\ket{0_{\text{\rom{1}}}} = \ket{0}$.

For the excitations of type \rom{2} consider the Hilbert space of spin-$\frac{1}{2}$ fermions that can have an additional flavour index $\alpha$, satisfying $\{c_{i\sigma}^{\alpha\dagger},  c^\beta_{j\sigma'} \}=\delta_{\alpha\beta}\delta_{\sigma\sigma'}\delta_{ij}$, where $\sigma, \sigma'\in\{\uparrow, \downarrow\}$ are spin indices.

Let $\tilde A_{\alpha\beta}$ be a $K\times K$ unitary matrix and $M_j = n_{j\uparrow} - n_{j\downarrow}$ - magnetization at site $j$.
The bilinear of type \rom{2}
\begin{align}
\label{eq:type2O}
\CO_j = \sum_{\alpha,\beta=1}^K c^{\alpha\dagger}_{j \downarrow} \tilde A_{\alpha\beta} c^\beta_{j \uparrow} \quad
\left[\CO_j, \CO_j^\dagger\right] = - M_j
\end{align}
creates "magnetic" excitations over the half-filled state with all spins down
$\ket{0_{\text{\rom{2}}}} = c^{1\dagger}_{j \downarrow} \cdots c^{K\dagger}_{j \downarrow}  \ket{0}$.

The operators $\CO_j$, $\CO_j^\dagger$ of either type and their commutator generate an SU$(2)$ algebra allowing for the following identifications
\begin{align}
\label{Jj}
J^{(j)}_+ \Leftrightarrow \CO_j,\quad  J^{(j)}_- \Leftrightarrow \CO^\dagger_j , \quad J^{(j)}_3 \Leftrightarrow \frac{1}{2}[\CO_j , \CO_j ^\dagger ]~. 
\end{align}
For two specific choices of matrix $A$ and single-flavour case this observation was made in Ref. \cite{yang1990so} and the two SU$(2)$ groups were identified as spin and pseudo-spin.

For both excitation types the maximum value of $J^{(j)}_3$ is limited by $K/2$ hence the largest representation is of "spin" $K/2$ meaning the operator $\CO_j^\dagger$ can be applied to its respective on-site vacuum the maximum of $K$ times. 
Furthermore, the operators $(\CO_j)^K$ for different $A$ coincide up to a constant factor because $(\CO_j)^K$ transforms the unique state with the minimum spin into the unique state with the maximum spin and is thus unique itself.

\emph{The O($N$)-invariant MBS}---Matrices $A_{\alpha\beta}$ and $\tilde{A}_{\alpha\beta}$ are site $j$-independent and the operator $\CO = \sum_j \CO_j$ is O$(N)$-symmetric,
where the O$(N)$ group, generated by $\sum_\alpha c_{i}^{\alpha\dagger} c^\alpha_{j}-c_{j}^{\alpha\dagger} c^\alpha_{i}$, 
acts on the site label of the fermionic operators and O$(N)$-invariance can be roughly thought of as independence on any relabelling of sites.

The wavefunctions
\begin{align}
\label{eq:genTowerWf}
\ket{\phi_n} = \frac{(\sum_i\CO^{\dagger}_i)^n \ket{0_\phi}}{P_N(n)}, \quad 0\le n\le NK,
\end{align}
with $P_N(n) = \sqrt{ \frac{ (NK)!n!}{ (NK-n)! } }$ and $\ket{0_\phi}=\ket{0_{\rom{1}/\rom{2}}}$  have symmetry of at least O($N$) while the full symmetry $G$, depending on $\CO^{\dagger}_j$ may be higher and would then contain O$(N)$ as a sub-group.

Due to O$(N)$-invariance any state from the subspace \eqref{eq:genTowerWf} is a many-body scar \cite{pakrouski2020GroupInvariantScars} for any Hamiltonian of the form 
\begin{align}
\label{eq:Hgeneral}
H=H_0 + OT,
\end{align}
where $O$ is an arbitrary operator (unrelated to the pairing operator $\CO$), $T$ is a generator of $G$. $H_0$ has to satisfy $[H_0, C^2_G]= W_c \cdot C^2_G$, where $W_c$ is some operator and $C^2_G$ is the quadratic Casimir of the group $G$.  In particular, this condition is satisfied by an $H_0$ that is $G$-invariant. Example Hamiltonian terms that can be used as $H_0$ or $T$ for single-orbital fermionic Hilbert space have been discussed in Ref. \cite{Pakrouski:2021jon}. 
The spectrum within the scar subspace spanned by the tower $\ket{\phi_n}$ \eqref{eq:genTowerWf} is determined solely by $H_0$ which takes the form of a $(NK+1)\times(NK+1)$ matrix.

Diagonal entries of this matrix can only come from the commutator $[\CO,\CO^\dagger]$ which are chemical potential/magnetization terms. We may introduce the restriction that for type-\rom{1} all the chemical potential terms $\mu_\alpha n_\alpha$ must have the same strength with $\mu$ thus independent of flavour $\alpha$. Similarly, for type-\rom{2} we may require that all the couplings in the terms of the form $B^\alpha_z (n^\alpha_{\uparrow}-n^\alpha_{\downarrow})$ are the same with $B^\alpha_z$ independent of $\alpha$.

The terms above and below diagonal come from $\CO$ and $\CO^\dagger$ respectively.
Any other matrix element would connect $\ket{n}$ with $\ket{m}$ where $|m-n|\ge2$ and can only be created by a Hamiltonian term that creates two or more excitations: $(\CO^\dagger)^p$ with $p\ge2$.
Any such operator would be at least 4-body and if we exclude such (unphysical and non-local since $\CO^\dagger$ includes the sum over all sites) operators then any valid O$(N)$-invariant $H_0$ within the scar subspace expands as a linear combination of the 3 SU$(2)$ generators: $[\CO,\CO^\dagger]$, $\CO$ and $\CO^\dagger$.

Thus
\begin{align}
\label{eq:H0generalSU2}
&H_0 = - \sum_j \mu  [\CO_j,\CO^\dagger_j] + \delta H_0 + C~,\\ 
&\delta H_0  = -\gamma \sum_j \left( e^{-i\theta} \CO_j + e^{i\theta} \CO^\dagger_j \right)
\label{eq:dH0general}
\end{align}
where $C$, $\mu,\gamma$ and $\theta$ are real numbers. For type \rom{1} $\mu$ plays the role of the chemical potential while for type \rom{2} of the $z$ magnetic field.

For type-\rom{1} the Hamiltonian $H_0$ \eqref{eq:H0generalSU2} coincides with the BCS mean-field Hamiltonian written down in real space (can be seen explicitly in the detailed examples discussion in the Appendix, for instance comparing eqs. \eqref{eq:H0etaInSpinorB},\eqref{eq:BCSHam} ). While it is typically thought of as an approximation to an interacting problem it is actually the exact Hamiltonian of strongly-interacting electrons within the scar subspace!

\emph{Solution for $H_0$}---Summing the local SU$(2)$ generators \eqref{Jj} over all sites $J_a \equiv \sum_j J_a^{(j)}$ we obtain a global SU$(2)$ algebra and the full Hilbert space decomposes into its irreducible representations. The largest representation has dimension $NK+1$ and the corresponding highest weight state is $\ket{0_\phi}$. 

Using eq. \eqref{Jj} we rewrite $H_0$ as
\begin{align}
\label{H00}
&H_0 = C + 2\sqrt{\gamma^2+\mu^2}H_z~,
\end{align}
where $H_{z} =-\frac{e^{-i\theta}J_+ + e^{i\theta}J_-}{2\cosh\tau}+ \tanh\tau J_3$, $\sinh \tau\equiv -\frac{\mu}{\gamma}$ and $C_{\rom{1}}=\mu K N$ and $C_{\rom{2}}=0$.

Consider a dimension $2h+1$ irreducible representation of SU$(2)$. A basis that diagonalizes $J_3$ is $|n\rangle$, i.e. $J_3|n\rangle = n|n\rangle$, where $n=-h, -h+1, \cdots, h-1, h$. In terms of this  basis, the largest eigenvalue of $J_1$ corresponds to the coherent state $e^{J_-}|h\rangle$, where $J_\pm = J_1\pm i J_2$. More generally, for any $z\equiv e^{\tau+i\theta}\in\mathbb C$, define (unnormalized) coherent state 
\begin{align}
\label{eq:genCoherentState}
|z_0 ) \equiv e^{z J_-}|h\rangle, 
\end{align}
satisfying the inner product 
\begin{align}\label{wz}
\quad (\bar w|z_0) = (1+\bar w z)^{2h}, \quad \bar w \equiv w^*.
\end{align}
The action of the SU$(2)$ algebra on this coherent state is given by 
\begin{align}
J_+|z_0) = &(2h z- z^2\partial_z)|z_0), \quad J_- |z_0) = \partial_z|z_0), \nonumber\\
& J_3 |z_0) = (h-z\partial_z)|z_0)~.
\end{align} 
For any fixed $z$, the state  $|z_0)$ is the (lowest) eigenvalue $-h$ eigenstate of $H_z$.
This can be checked by first imposing the cancellation of the $\partial_z$ terms and then tuning the overall normalization such that it is equal to $-h$.
The full spectrum of $H_{z} $ is the same as that of $J_3$: $E_n = n\in\{-h, -h+1, \cdots, h-1, h\}$.

The excited states 
\begin{align}
\label{eq:zn}
\ket{z_n} = \frac{(\sum_i\CO^{\gamma\dagger}_i)^n \ket{z_0}}{P_N(n)}, \quad 0\le n\le NK,
\end{align}
have the form \eqref{eq:genTowerWf} where $\ket{0_\phi}$ is replaced with $\ket{z_0}$ and $\CO^\dagger$ with
\begin{align}
\label{eq:gammaRaisingOpGeneral}
\CO^{\gamma\dagger} = \frac{\bar z J_+- \bar z^{-1} J_-}{2\cosh\tau}+ \frac{J_3}{\cosh\tau}~
\end{align}

and form a full basis in the dimension $2h+1$ irreducible representation we started with.

The ground state energy of $H_0$ \eqref{H00} is $E_{0} =C -K N\sqrt{\gamma^2+\mu^2}$
and the corresponding product ground state
\begin{align}
\label{eq:gsGeneralExp}
|z_0\rangle  
= N_z \prod_j e^{ \frac{v}{u} \CO_j^\dagger}\ket{0_\phi},\quad N_z = u^{NK}~,
\end{align}
where 
\begin{align}
\small
\label{eq:uvXdef}
 \quad u=\sqrt{\frac{1+\frac{\mu}{E}}{2}}, \quad v=e^{i\theta}\sqrt{\frac{1-\frac{\mu}{E}}{2}}, \quad
E = \sqrt{\gamma^2+\mu^2}.
\end{align}
\normalsize

The wavefunction \eqref{eq:gsGeneralExp} coincides with the BCS wavefunction \cite{TinkhamBookIntroSC} where its coefficients $u$ and $v$ {\it are independent of the position on the lattice}.

For multi-flavour fermions with $K\ge2$ the expansion in \eqref{eq:gsGeneralExp} contains higher (up to $K$) powers of $ \CO_j^\dagger$ 
\begin{align}
\label{eq:gsGeneral}
|z_0\rangle =  N_z \prod_j \left( 1 + \frac{v}{u}\CO_j^\dagger + \frac{ (\frac{v}{u}\CO_j^\dagger)^2 }{2}  +\dots+  \frac{( \frac{v}{u}\CO_j^\dagger)^K }{K!} \right)\ket{0_\phi}
\end{align}
where terms $(\CO_j^\dagger)^m$ may be interpreted as clustering of $2m$ fermions by analogy with pairing for $m=1$. We expect that higher-$m$ clustering dominates for $\frac{v}{u}>1$ equivalent to $\mu<0$ for $\theta=0$.

The states $\ket{z_n}$ span the same subspace as $\ket{\phi_n}$ \eqref{eq:genTowerWf} and are equidistant in energy ($E_{n} =E_0  + n 2E$) MBS for a broad class of Hamiltonians of the form \eqref{eq:Hgeneral}. Any initial state formed as their linear superposition will return to itself in time evolution with the revivals period of $\frac{\pi}{\sqrt{\gamma^2+\mu^2}}$ \cite{pakrouski2020GroupInvariantScars}.

The filling factor for type \rom{1}  is $\langle z_n |\nu|z_n\rangle  =\frac{1}{2} + \frac{\mu}{\sqrt{\mu^2+\gamma^2}}(\frac{n}{NK}-\frac{1}{2})$
and the total $z$ axis magnetization for type-\rom{2} is
$\langle z_n |M|z_n\rangle = \frac{\mu}{\sqrt{\mu^2+\gamma^2}}(2n-NK)$.

In Appendix we demonstrate that some parts of the solution obtained here could also be derived for any fixed definition of the $\CO$ operator using a canonical Bogoliubov transformation. 
The raising operator $\CO^{\gamma\dagger}$ \eqref{eq:gammaRaisingOpGeneral} in this language is constructed according to \eqref{eq:type1O} or \eqref{eq:type2O} where the original fermionic operators are replaced with the Bogoliubov-transformed ones.

The relevance of the coherent states such as $\ket{z_0}$ \eqref{eq:genCoherentState} for Hubbard-like models was noticed long ago \cite{CohStatesRadcliffe1971,CohStatesOutOfEtaPairingStates1998}. The derivation of the states $\ket{z_n}$ is of course not the main contribution of our work - rather demonstrating that these states are O$(N)$-invariant MBS. Nevertheless we note that to our knowledge we for the first time show here that the tower of states $\ket{z_n}$ forms an alternative (to the generalization of the $\eta$ pairing states $\ket{\phi_n}$ \eqref{eq:genTowerWf}) full basis within the O$(N)$-invariant subspace.

\emph{Correlations and the global energy minimum}---Due to the O$(N)$ invariance, the one-point function $\braket{\CO^\dagger_j}$, measured in any state from the subspace spanned by $|z_n\rangle $ \eqref{eq:gsGeneralExp} does not depend on the measurement point $j$ \cite{pakrouski2020GroupInvariantScars} and the two-point function $\braket{\CO^{\dagger}_i \CO_j}$ does not depend on the distance between the two points $i$ and $j$. The latter property by definition \cite{etaPairingYang89} means that any state from the scar subspace has the off-diagonal long-range order of type $\CO^{\dagger}_j$.

The 1- and 2-point functions in MBS are given by (see Appendix for derivation)
\begin{align}
\label{eq:1ptFunInSPGSK1}
\braket{z_n|\CO^\dagger_j|z_n}  = \frac{  e^{-i\theta} } {\sqrt{1+(\frac{\mu}{\gamma})^2}} \left( \frac{K}{2}-\frac{n}{N} \right)
\end{align}
and
\begin{align}
\label{eq:2ptFunInZInTermsOfMu}
\braket{z_n|\CO^\dagger_j \CO_i|z_n}  = \frac{ 4Kn(KN-n) + \frac{K\left( 6n^2 - 6KnN + KN(KN-1) \right)}{1 + (\frac{\mu}{\gamma})^2 } }{4N(KN-1)}.
\end{align}
The 1-point functions \eqref{eq:1ptFunInSPGSK1} decrease monotonically with $\mu$ and we show in the Appendix that it is bound (over the full Hilbert space) from above by $\frac{K}{2}$. This bound is exactly saturated by the state $\ket{z_0}$ for $\mu=0$ (value is system-size $N$-independent).
This means that for any Hamiltonian of the form \eqref{eq:Hgeneral}, there exists a large enough $\gamma_c$ (dependent on the strength of $OT$) such that the scar state $|z_0\rangle$ \eqref{eq:gsGeneral} becomes the ground state for $\mu=0$.

The average of the 2-point function (ODLRO) over all states in the scar subspace is basis-independent

\begin{align}
\label{eq:2ptFunAvgOverScarSubsp}
\frac{ \sum_{n=0}^{NK} \braket{z_n|\CO^\dagger_j \CO_i|z_n}}{NK+1} = \frac{K^2}{6}.
\end{align}

\emph{Admissible $OT$ terms in the Hamiltonian} --- depend on the full symmetry group of the states $\ket{\phi_n}$ \eqref{eq:genTowerWf} which in turn depends on the specific pairing matrix $A_{\alpha\beta}$ (or $\tilde A_{\alpha\beta}$ for type-$\rom{2}$). By construction, O$(N)$ is always a sub-group of the full symmetry. Therefore the generators of O$(N)$, the hopping terms $T^h_{ij} = i\sum_{p,\alpha} \left(c^{p,\dagger}_{i\alpha}\, c^{p}_{j\alpha}-\text{H.c.}\right)$ can always be used as $T$ in the $OT$ part of the Hamiltonian.

The full symmetry group of the states $\ket{\phi_n}$ \eqref{eq:genTowerWf} for any specific pairing is likely to be larger than O$(N)$ and needs to be worked out on the case-by-case basis to identify additional (to the hopping) generators that can be used as $T$.

\emph{Revivals and sensitivity to perturbations} --- We have shown so far that the states $\ket{\phi_n}$ \eqref{eq:genTowerWf} are exact MBS for any Hamiltonian of the form \eqref{eq:Hgeneral} with the admissible $T$ discussed above, $O$ - arbitrary operator and $H_0$ in eq. \eqref{eq:H0generalSU2}. For any such Hamiltonian the revivals of the initial wavefunction from the MBS subspace are exact and infinite. The effect of the perturbations deviating from the form \eqref{eq:Hgeneral} depends on the specific pairing matrix, must therefore be studied on a case-by-case basis and is not the subject of the present work aiming to outline the mechanism of MBS for general pairing. The effects of perturbations relevant for our first example below have been studied in detail in Ref. \cite{KolbStabilityPRXQuantum2023}.

\emph{For the first example} --- (worked out in detail in the Appendix) consider K=1, spin-1/2 fermions on a 2D square lattice and the standard Hubbard Hamiltonian with the additional pairing potential $\delta H_0$
\begin{align}
\label{eq:HimHubMain}
H^{H} = & \mu_H \sum_{i=1}^N (n_{i\uparrow}+ n_{i\downarrow})  + \sum_{\braket{kj}} T_{kj} \\ \nonumber
+& U \sum_j n_{j\downarrow}n_{j\uparrow}  +\delta H_0,
\end{align}
where $T_{kj} = t \sum_{\sigma p} c^{p\dagger}_{j\sigma} c^p_{k\sigma} +h.c.$
Three families of scar states of the form \eqref{eq:genTowerWf} may exist in this model \cite{pakrouski2020GroupInvariantScars,Pakrouski:2021jon} for $\gamma=0$ defined by three different $\CO_j^{\dagger}$ 
\begin{align}
\label{eq:1bandPairingOdagMain}
\eta_j^\dagger  = c^\dagger_{j\uparrow} c^\dagger_{j\downarrow};  \quad
{\eta '}_j^{\dagger}  =   e^{i\pi j} c^\dagger_{j\uparrow} c^\dagger_{j\downarrow}; \quad
{\zeta}_j^{\dagger}  = c^\dagger_{j\uparrow} c_{j\downarrow}.
\end{align}

The full catalog of Hamiltonian terms supporting these three families of states as scars can be found in Ref. \cite{Pakrouski:2021jon}.

For real $t$ and $\CO_j^{\dagger}={\eta '}_j^{\dagger}$ the scar states \eqref{eq:genTowerWf} are the well-known \cite{etaPairingYang89} eta-pairing states $\ket{n^\eta}'$ defined on any bipartite lattice. These states' symmetry does include O$(N)$ as a subgroup as was shown in Ref. \cite{Pakrouski:2021jon} (see Fig. 1 and discussion around eqs. (19-21)). Although these states have an alternating $\pm$ sign between the two sublattices, on a bipartite lattice one can get rid of this sign by a unitary transformation that turns the $\ket{n^\eta}'$ states into the site-uniform $\ket{n^\eta}$. While both families of states have O$(N)$ symmetry they correspond to different implementations of O$(N)$, therefore Ref. \cite{Pakrouski:2021jon} denoted one of them as O$(N)'$

\begin{figure}[htp!]
	\begin{center}
				\includegraphics[width=0.46\columnwidth]{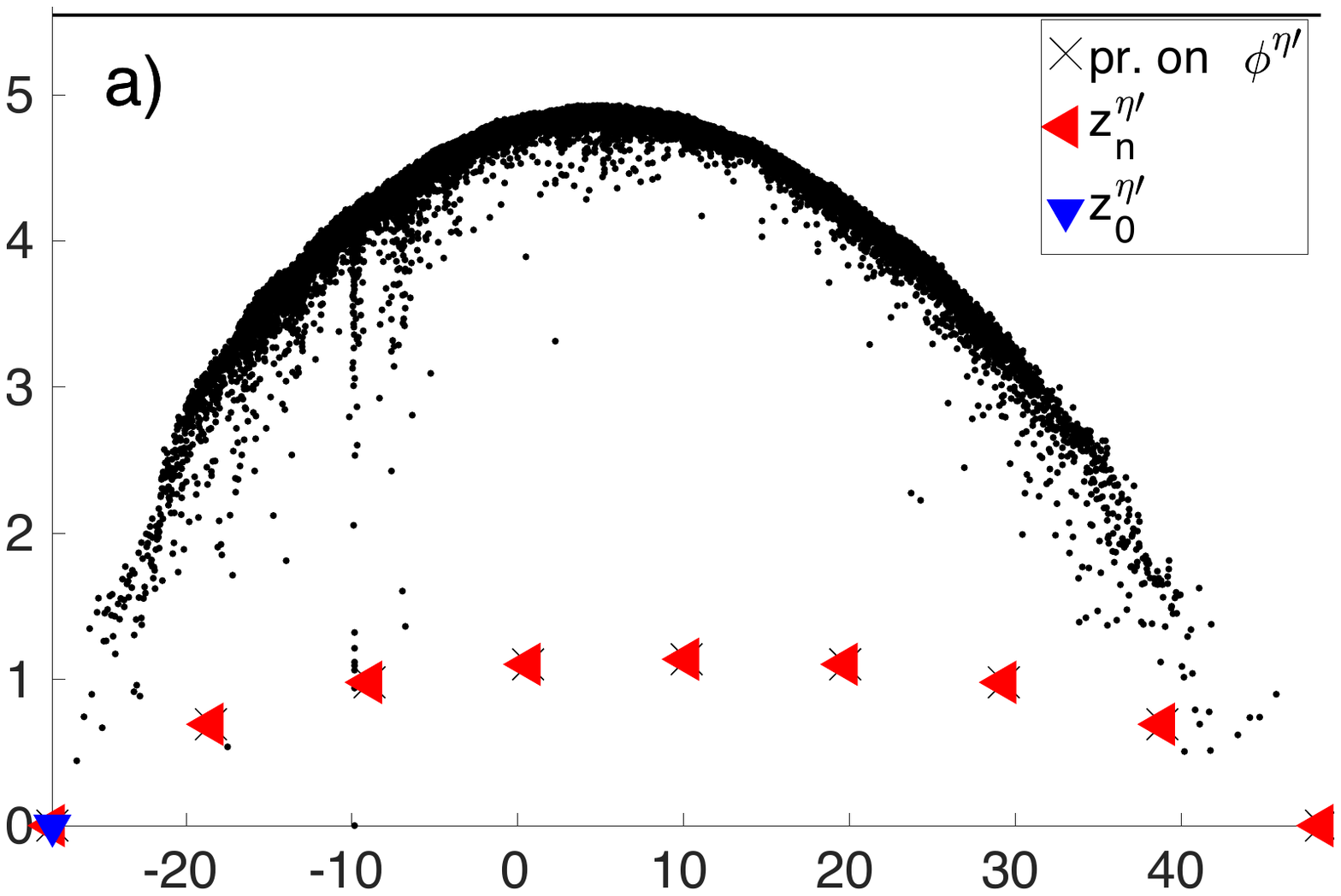}	
				\includegraphics[width=0.46\columnwidth]{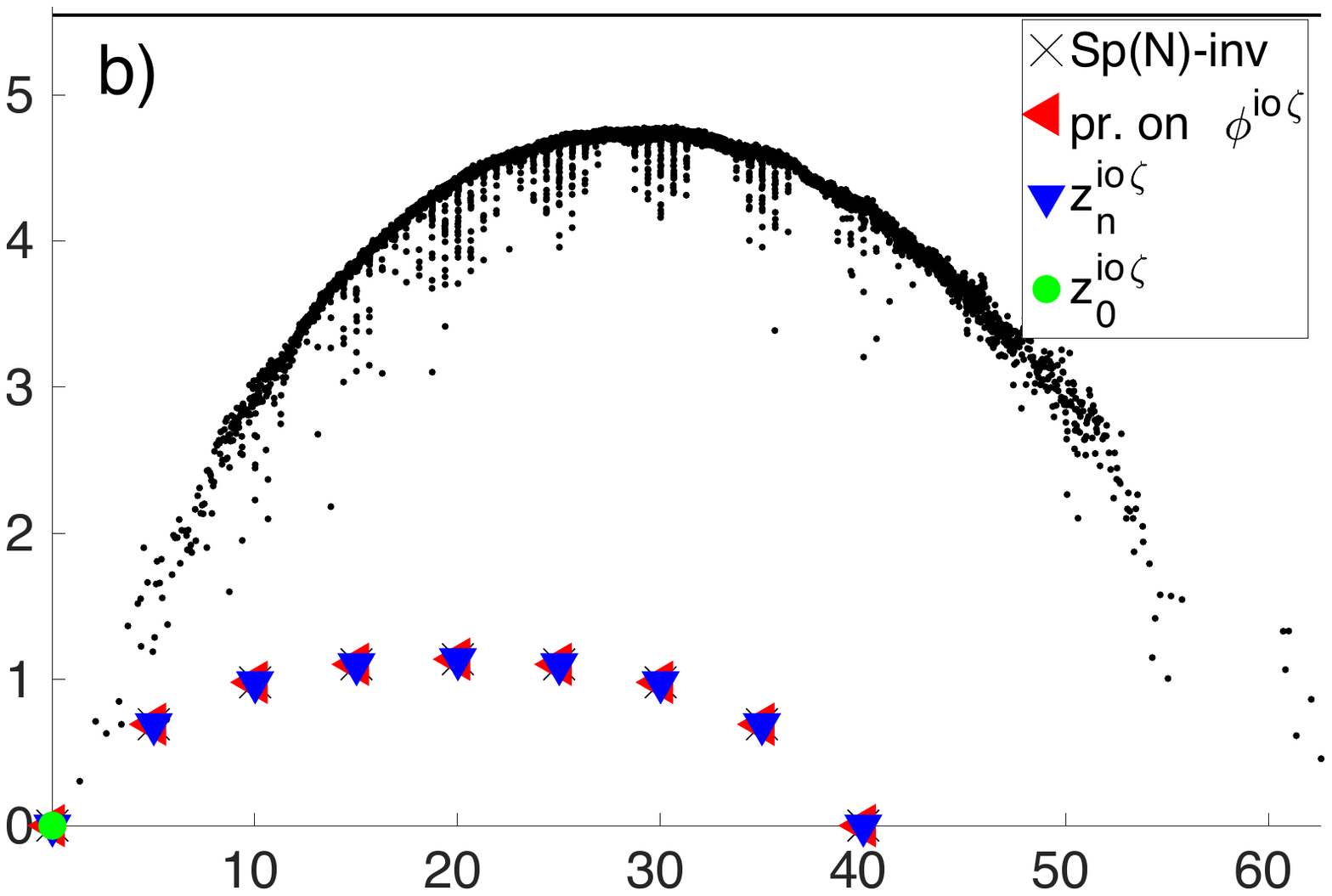}
				\includegraphics[width=0.49\columnwidth]{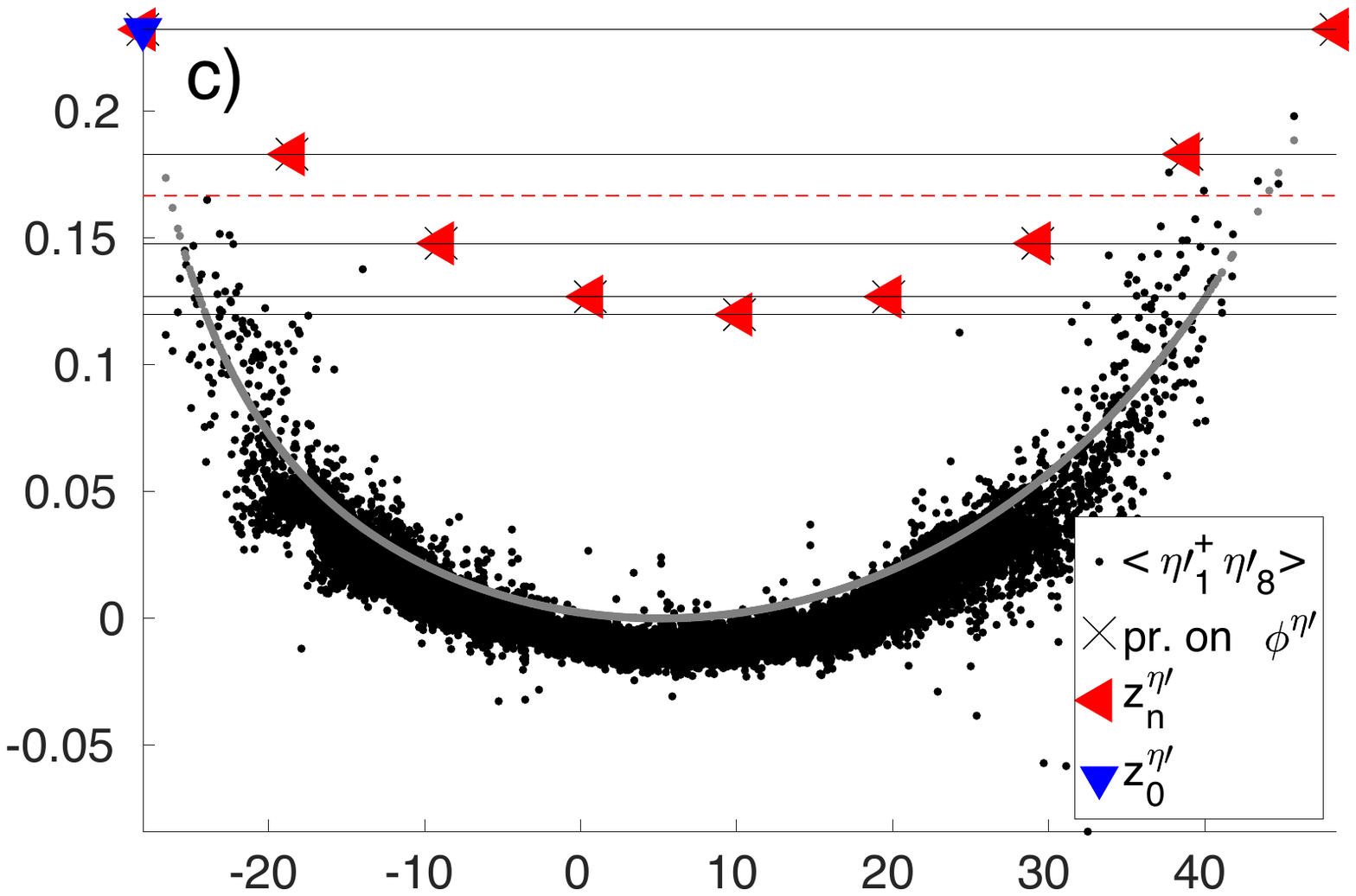}
				\includegraphics[width=0.49\columnwidth]{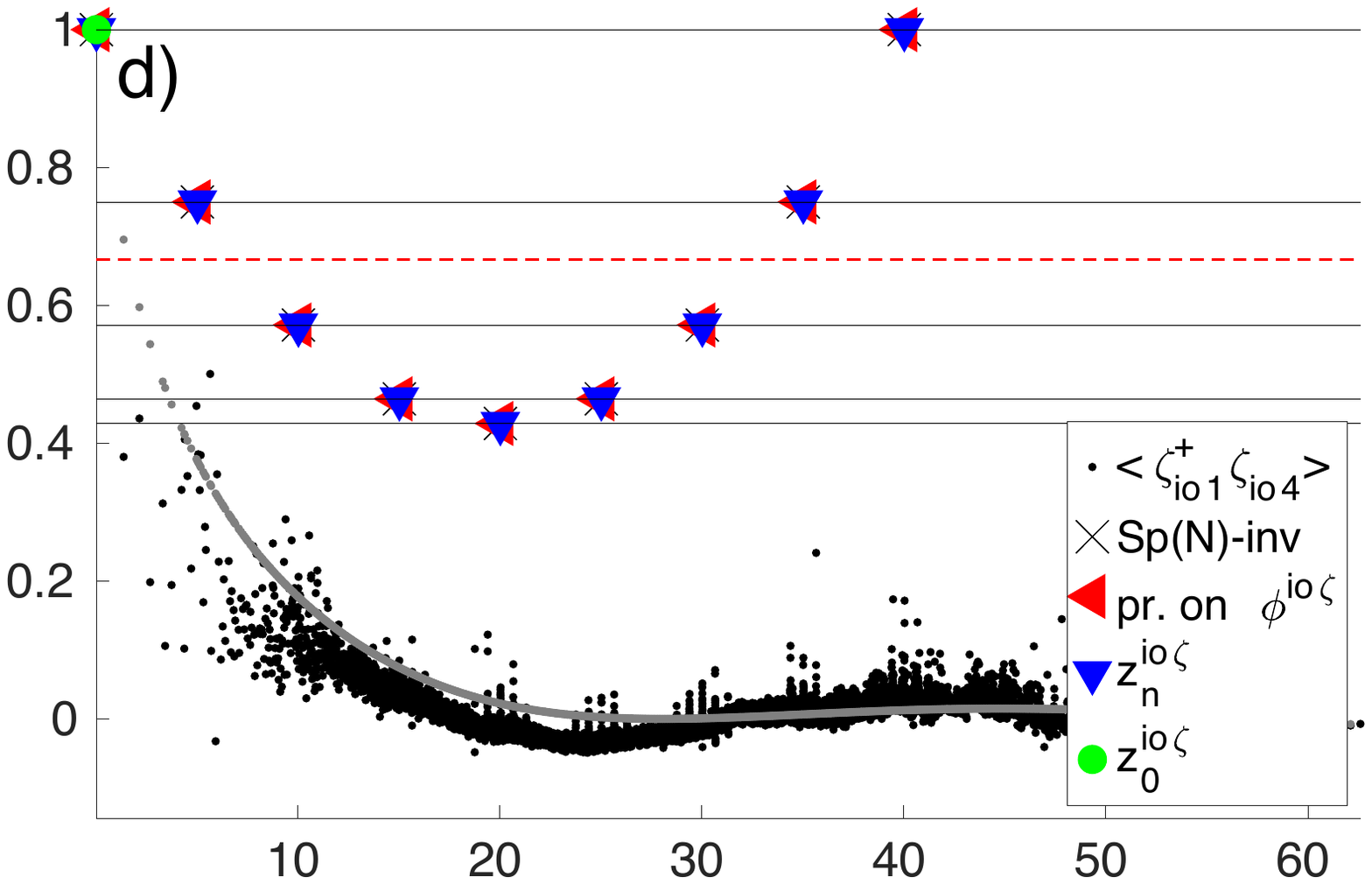}
				\includegraphics[width=0.49\columnwidth]{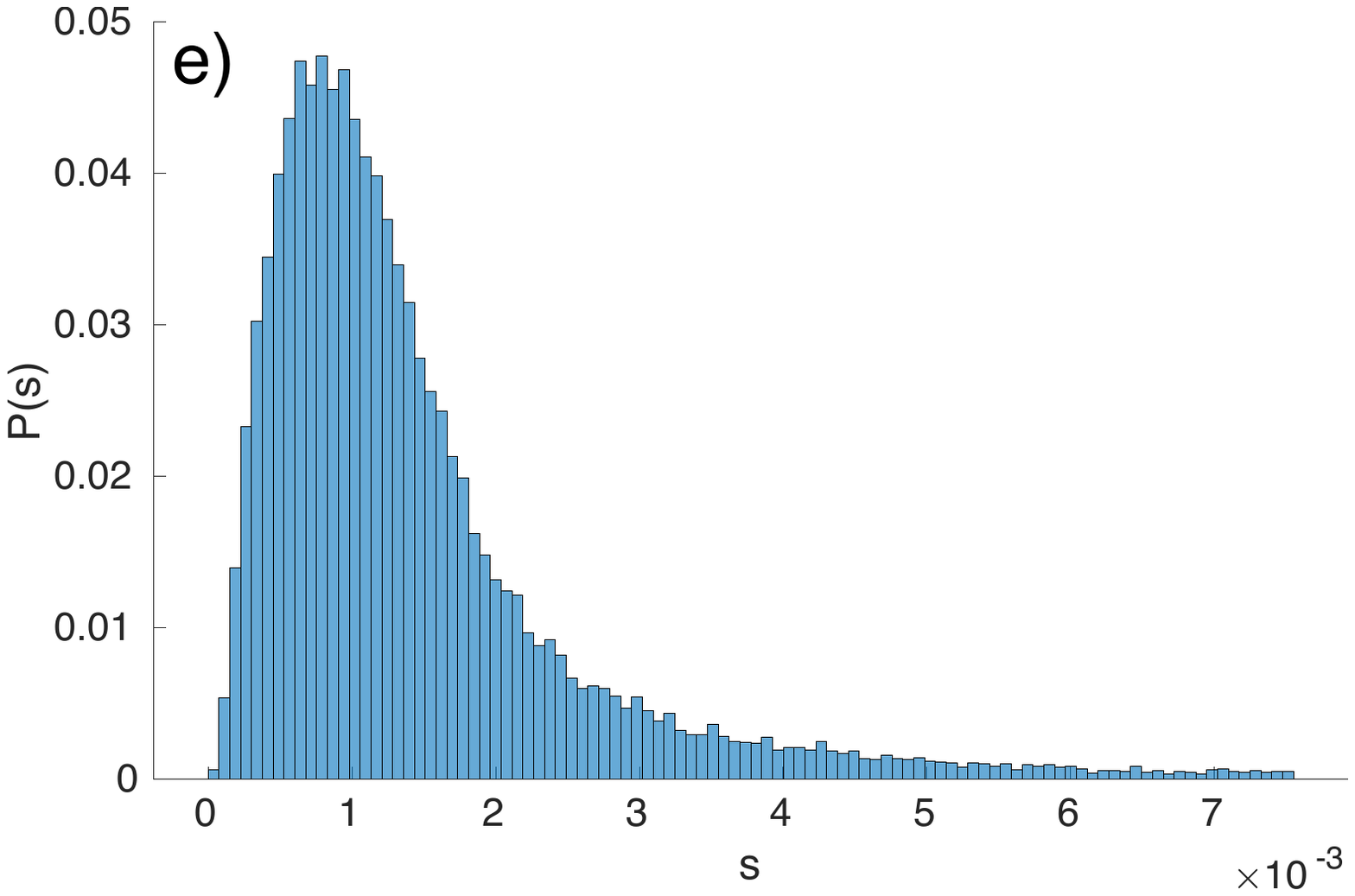}
				\includegraphics[width=0.49\columnwidth]{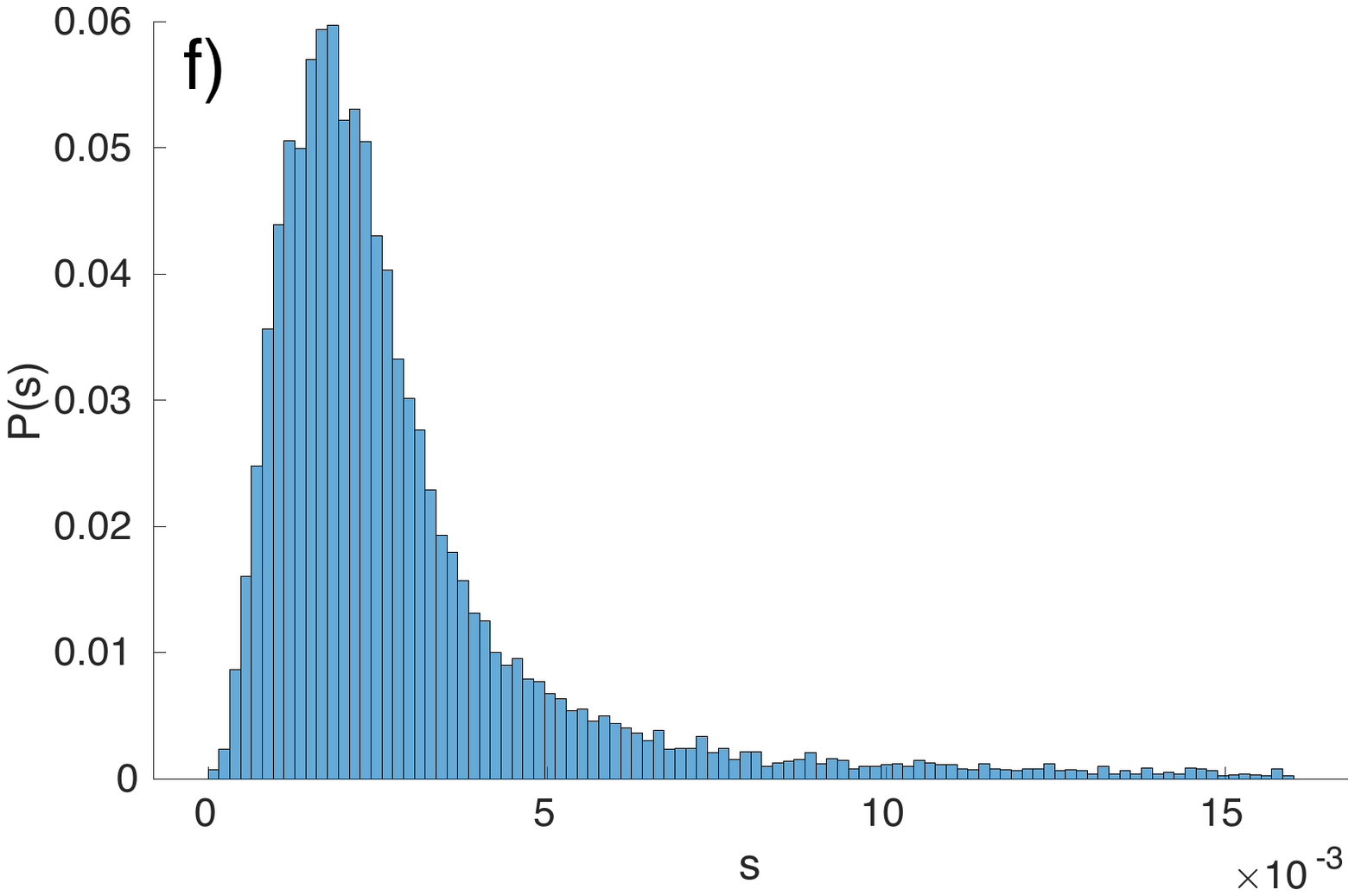}
				
	\end{center}
\caption{\label{fig:numericsEtaPandibMagnMain} Numerical results for (a),c),e),left column) $K=1$, 2D real-hopping Hubbard model with o.b.c., $U=5.01$, $\mu=-1.2295$, $\theta=\pi/7$, $N_x\times N_y=2\times4$ and $\gamma=4.61$ and (b),d),f),right column) for $K=2$, $\CO$ from \eqref{eq:ibZetaOdagMain} and the generalized Hubbard Hamiltonian \eqref{eq:ibZetaHMain}. 1D, p.b.c., $N=4$, $t=i$, $B^p_z=\{0.3147; -0.3147\}$, $\theta=\pi/7$, $U=5.01$, $\gamma=2.5$, $r_j=\{0.132, 0.17, -0.157, 0.174\}$, fixed total particle number (half-filling). a),b) Entanglement entropy for a cut separating the system into 2 equal pieces (cut horizontally in the 2D case). All the $\ket{z_n}$ states have unity projection on the subspace spanned by $\ket{\phi_n}$. c,d) ODLRO $\braket{\CO^\dagger_i\CO_j}$ with $i=1$ and $j=N$ - the two sites with largest separation in our lattice. Dashed horizontal line indicates the average over the scar subspace \eqref{eq:2ptFunAvgOverScarSubsp}. The thermal average over all eigenstates is indicated in c) and d) with a grey line. e),f) Histogram of all the energy gaps in the spectrum, excluding the MBS states. The horizontal axis in panels a),b),c),d) is energy.
}
\end{figure}

In presence of the pairing potential, for $\gamma>0$, the lowest energy scar is the BCS wavefunction
\begin{align}
\label{eq:etaPspgsMain}
|z^\eta_0\rangle =  N_z \prod_j \left( 1 + \frac{e^{i\theta}}{\frac{\mu}{\gamma} + \sqrt{\frac{\mu^2}{\gamma^2} + 1} }  e^{i\pi j} c^\dagger_{j\uparrow} c^\dagger_{j\downarrow} \right)\ket{0},
\end{align}
where $\mu=\frac{U}{2}+\mu_H$.
This state (originally discussed in \cite{CohStatesOutOfEtaPairingStates1998}) and its excitations are a linear combination of the states in the original tower \eqref{eq:genTowerWf} which we confirm numerically in the left column of Fig. \ref{fig:numericsEtaPandibMagnMain}. 
We thus establish that the BCS-like wavefunction $|z^\eta_0\rangle$ is a linear combination of the eta-pairing states!

In both cases shown in Fig. \ref{fig:numericsEtaPandibMagnMain} $\gamma$ was chosen to be about the smallest by absolute value for which the MBS state becomes the ground state for the given system size. 
For this reason the generic states appear close to the MBS ground state.
Nevertheless, the tower of states $\ket{z_n}$ remains MBS which in particular means this subspace remains decoupled from the generic states and all its special properties such as strong correlations are preserved. 
Furthermore no transitions or excitations between the MBS and generic states are possible.
The $\ket{n^\zeta}$ states (see \cite{Pakrouski:2021jon} and Appendix) are degenerate with energy $\mu N\approx-9.84$ for the Hamiltonian \eqref{eq:HimHubMain} and although not the focus of this example can be found as a degenerate group in Fig \ref{fig:numericsEtaPandibMagnMain}a).

In numerics used to obtain Fig. \ref{fig:numericsEtaPandibMagnMain}, we add to the Hamiltonian an auxiliary, symmetry-breaking $OT$ term
\begin{gather}
\label{eq:TOT2}
\sum_l O_l T_l =  \sum_{\braket{ij},p} T_{ij} O_{ip} T_{ij}
\end{gather}
with $O_{ip} =   \left( r^{(1)i,p}_{\sigma, \sigma'}c^{p\dagger}_{i \sigma} c^{(p+1)}_{i \sigma'} + r_{\sigma, \sigma'}^{(2)i,p}c^{p\dagger}_{i \sigma} c^{(p+1)\dagger}_{i\sigma'} + {\rm h.c.} \right)$ and $r^{(1/2)i,p}_{\sigma, \sigma'}$ - real random numbers, between 0 and 1. It has no effect on the MBS subspace and ensures Hamiltonian has no remaining symmetries besides total fermion number parity.

The e) and f) panels of Fig. \ref{fig:numericsEtaPandibMagnMain} show the distribution of the energy gaps in the bulk of the spectrum (excluding the MBS states). Near-zero energy gaps are almost absent in agreement with the absence of any remaining symmetries, indicating the level repulsion throughout the spectrum. In panels c),d) we observe that the two-point function measured in the group-invariant states (and the average over these states) substantially deviates from the thermal average indicated by the grey datapoints. Similar observation can be made about the entanglement entropy in panels a),b). We conclude that the group-invariant states do violate the Eigenstate Thermalisation Hypothesis and may thus be called many-body scars.

\emph{For the second example} --- consider a model with $K=2$ orbitals labeled $A$ and $B$ with type-\rom{2}
\begin{align}
\label{eq:ibZetaOdagMain}
\CO^{\dagger}_j = c^{A\dagger}_{j\uparrow}c^B_{j\downarrow} + c^{B\dagger}_{j\uparrow} c^A_{j\downarrow}
\end{align}
that couples the spin and orbital degrees of freedom.

We start from a generalized Hubbard Hamiltonian in external magnetic field and including spin-orbit coupling

\begin{align}
\label{eq:ibZetaHMain}
H^{B} = &\sum_{i=1}^N \sum_{p=A}^B B^p_z (n^p_{i\uparrow}- n^p_{i\downarrow})  +  \sum_{\braket{kj}} T_{kj} \\ \nonumber
+ &U \sum_j H_j^{\zeta Hub} + H_{SO} + \delta H_0,
\end{align}
where the hopping amplitude $t$ is imaginary and $H_j^{\zeta Hub} = \frac{1}{2} (n_j^2 - n_j)$
which in the single-orbital case reduces to the standard Hubbard interaction. It is also a special case of more general Hubbard interaction \cite{typicalUSCHamZinkl2021} that includes intra-orbital, inter-orbital and Hund's rule couplings with fixed relative strength. $H_{SO}$ is the on-site spin-orbit coupling relevant for certain well-studied 2-orbital materials \cite{typicalUSCHamZinkl2021}
 \begin{gather}
\label{eq:HsoCorrectWithSigma}
H_{SO} =  -\lambda i \sum_{j\alpha\beta} r_j (\sigma^z)_{\alpha\beta} c^{A\dagger}_{j\alpha} c^B_{j\beta} + \mathrm{H.c.},
\end{gather}
where $r_j$ are real numbers.

In terms of the form \eqref{eq:Hgeneral} the $H_0$ part of the Hamiltonian \eqref{eq:ibZetaHMain} only includes the $B_z$ and $\delta H_0$ terms leading to $\mu=0.5(B^A_z+B^B_z)$ in the general solution.
The lowest energy scar state is given by
\begin{align}
\label{eq:gsIBZetaMain}
|z^{io\zeta}_0\rangle =  N_z \prod_j \left( 1 + \frac{v}{u}\CO_j^\dagger + \frac{ (\frac{v}{u}\CO_j^\dagger)^2 }{2}  \right)\ket{0_{\rom{2}}}.
\end{align}

The numerical results in the right column of Fig. \ref{fig:numericsEtaPandibMagnMain} (also here the auxiliary term \eqref{eq:TOT2} is added) confirm the states $|z^{io\zeta}_n\rangle$ are many body scars with strong correlations of type $\CO_j^\dagger$. These states have at least Sp$(N)$ symmetry (which includes O$(N)$ as a subgroup).

The total magnetization in the scar states is proportional to the effective chemical potential $\mu$ and is zero for $B^A_z=-B^B_z$ as in Fig. \ref{fig:numericsEtaPandibMagnMain}. Thus the MBS at the same time have zero total magnetization and large expectation values for the inter-orbital magnetic excitation creation operator $\braket{\CO^{\dagger}_j}$ and for the two-point function $\braket{\CO^{\dagger}_j\CO_i}$ independent of the site positions $i,j$. The ODLRO $\braket{\CO^{\dagger}_j\CO_i}$ in the scar subspace is significantly larger (factor of 27 on average) than in generic states.

For simplicity Hamiltonian \eqref{eq:ibZetaHMain} has uniform, site-independent hopping strength. However, every individual hopping term annihilates the MBS subspace exactly and the same is true for every individual spin-orbit term. Therefore the dynamics within the MBS subspace (and the decoupling of this subspace) is completely insensitive to perturbations that change the strength of hopping or spin-orbit.

\emph{Discussion}---The solution presented here does not materially depend on the matrices $A$ \eqref{eq:type1O} and $\tilde{A}$ \eqref{eq:type2O} that specify the particular type of pairing. Therefore our conclusions hold for any unitary pairing including unconventional! It would be interesting to extend our approach to pairing operators that act on more than one site.

As discussed following eq. \eqref{eq:2ptFunAvgOverScarSubsp} the off-diagonal long-range order averaged over the scar subspace \eqref{eq:genTowerWf} is independent of the Hamiltonian parameters and of the system size! It remains large with or without the pairing potential $\delta H_0$ added to the Hamiltonian.

Within the scar subspace the dynamics is fully governed by $H_0$ \eqref{eq:H0generalSU2} which coincides with the mean-field Hamiltonian. This presents an alternative view on the latter as the exact Hamiltonian within the scar subspace rather than an approximation to an interacting model. Its solution is the BCS-like wavefunction \eqref{eq:genCoherentState} and excitations \eqref{eq:zn} that are unaffected by a variety of $OT$ perturbations. We therefore demonstrate the exceptional stability of the mean-field solution to a broad class of interactions. BCS-like wavefunction will remain the ground state as long as these perturbations are small.

Since the state $|z_0\rangle$ has the maximum $\braket{\CO^{\dagger}_i}$ over the full Hilbert space for $\mu=0$ we can always, in principle, make it the ground state by adding strong enough $\delta H_0$. Because $\delta H_0$ is strictly local and relatively simple this provides the first feasible protocol to initialize a fermionic system to a scar state in (a quantum simulator) experiment, although arbitrarily adjusting the strength of $\delta H_0$ may in general be challenging. For type-\rom{1} the condition $\mu=\mu_H+U/2=0$ can be achieved by tuning the chemical potential $\mu_H$ using a global electric gate. For type-\rom{2} - by adjusting the magnetic field.

Let $\gamma_c$ be such that the ground state is in the scar subspace for $\gamma>\gamma_c$ and in the remainder of the Hilbert space for $\gamma<\gamma_c$. Consider zero temperature, $\gamma>\gamma_c$ and assume the initial state of the system is within the scar subspace. Slow tuning of $\gamma$ will not lead to a phase transition to a state outside the scar subspace even as we pass $\gamma_c$ because due to the scar subspace decoupling (which is the source of ergodicity breaking) there are no matrix elements that could transform the wavefunction into a "would be" non-scar ground state. Alternatively, from the same starting point we could keep $\gamma$ fixed and increase temperature. The system again can not leave the scar subspace and will turn into a linear combination of MBS each possessing strong ODLRO. (Same conclusions would apply for exact sectors of a symmetry while there is no symmetry of the full $H$ in our system that would trivially isolate MBS.) It would be interesting to study how the above predictions change if the Hamiltonian slightly deviates \cite{KolbStabilityPRXQuantum2023} from the $H_0+OT$ form.

The numerics presented confirms in small systems the key analytical results because the MBS wavefunctions are exactly identified among the numerical eigenstates with all their properties as predicted.
We demonstrated analytically that the states $\ket{\phi_n}$ (for $\gamma=0$) and $|z_n\rangle$ (for $\gamma\ne0$) are MBS for suitable Hamiltonians on any lattice and for any system size.
The value of $\gamma_c$ that is large enough for a scar to become the ground state is system size-dependent and we make no predictions about it besides arguing that such $\gamma_c$ exists.
$\gamma_c$ depends on the particular terms in the $OT$ part of the Hamiltonian. Its system-size dependence is an interesting question for a separate future numerical study focusing on a specific Hamiltonian.

\emph{Acknowledgements}---We thank Manfred Sigrist for many illuminating discussions and acknowledge useful exchanges with Kirill Samokhin, Yuto Shibata and Ilaria Maccari.
We also thank Igor Klebanov and Fedor Popov for the collaboration on the several related publications in preparation.
The simulations presented in this work were performed on computational resources managed and supported by Princeton's Institute for Computational Science $\&$ Engineering and OIT Research Computing.

\appendix

\section{More details on the general solution}

The 1- and 2-point expectation values in the coherent state $\ket{z}$ and the excited states $\ket{z_n}$ are given by 
\begin{align}
\label{eq:1ptFunInZK1}
\braket{z_n|J^{j}_-|z_n}  =\frac{ e^{-i\theta} } {\cosh\tau} \left( \frac{K}{2}-\frac{n}{N} \right)
\end{align}
\begin{align}
\label{eq:2ptFunInZ}
\braket{z_n|J^{i}_+J^{j}_-|z_n}  =\frac{K^2}{4\cosh^2\tau}+\frac{nK(KN-n)}{2N(KN-1)}\frac{\cosh(2\tau)-2}{\cosh^2\tau}.
\end{align}

We also find 
\begin{align}
\label{eq:totQInZ}
 \langle z_n| J_3|z_n\rangle =  \tanh\tau\left (n- \frac{NK}{2}\right)
\end{align}

As we argued the on-site SU(2) formed by the operators $\CO_j$ and $\CO^\dagger_j$ has the largest representation with spin $K/2$ leading to the maximum value of $K/2$ for any of the three generators $J_\alpha$ measured in any state in the Hilbert space. 

Therefore we have 
\begin{align}
\label{eq:boundOnOffDiagH0}
\CO_j + \CO_j^\dagger = J_+^{j} + J_-^{j} = 2J_x^{j}\le K
\end{align}

As a consequence the bound (over all states in Hilbert space) on the value of the 1-point function is
\begin{align}
\label{eq:bound1p}
Re[\braket{\CO^\dagger_j}] =\frac{1}{2} \braket{ \CO_j + \CO^\dagger_j } = \frac{K}{2}.
\end{align}

The bound for the two-point function in any product state is
\begin{align}
\label{eq:bound2p}
\braket{\CO^\dagger_i\CO_j} = \braket{\CO^\dagger_i}\braket{\CO_j} \le \frac{K^2}{4}.
\end{align}

\section{Example 1: 2D single-orbital Hubbard model \label{sec:1orbitalHubbard}}

Consider the Hilbert space of spin-1/2 fermions without any additional flavours (K=1 for both types) on a 2D square lattice.

The starting point is the standard Hubbard Hamiltonian
\begin{align}
\label{eq:HimHub}
H^{H} = \mu_H \sum_{i=1}^N (n_{i\uparrow}+ n_{i\downarrow})  + \sum_{\braket{kj}} T_{kj} + U \sum_j n_{j\downarrow}n_{j\uparrow}.
\end{align}

\subsection{MBS and $H_0+OT$ decomposition}

There are three families of scar states in this model \cite{pakrouski2020GroupInvariantScars,Pakrouski:2021jon} for purely imaginary ($\ket{n^\eta}$), purely real ($\ket{n^\eta}'$) or complex ($\ket{n^\zeta}$) hopping amplitude $t$. Each of the three families is constructed according to eq. \eqref{eq:genTowerWf}
with the operator $\CO_j^{\dagger}$  is given by
\begin{align}
\label{eq:1bandPairingOdag}
\eta_j^\dagger  = \frac{1}{2} \left( c^\dagger_{j\uparrow} c^\dagger_{j\downarrow} - c^\dagger_{j\downarrow} c^\dagger_{j\uparrow} \right) \\ \nonumber
{\eta '}_j^{\dagger}  =  \frac{1}{2} e^{i\pi j} \left( c^\dagger_{j\uparrow} c^\dagger_{j\downarrow} - c^\dagger_{j\downarrow} c^\dagger_{j\uparrow} \right)\\ \nonumber
{\zeta}_j^{\dagger}  = c^\dagger_{j\uparrow} c_{j\downarrow},
\end{align}
with the excitations of type-\rom{1} in the first two cases and of type-\rom{2} for the third case.

As shown in Ref. \cite{Pakrouski:2021jon} the $H_0+OT$ decomposition of the Hubbard model for the $\eta$ and $\eta'$ states reads\\
\begin{gather}
H^{H} =H^{H\eta}_0 +  \sum_{\braket{kj}} T_{kj}  - \frac{U}{2} \sum\limits_i M_i^2,
\label{eq:HubDecompEta}
\end{gather}
where
\begin{gather}
H^{H\eta}_0 =n \left(\frac{U}{2} +\mu_H \right)
\label{eq:HubDecompEta}
\end{gather}
with $n$ total particle number operator and the remaining ($OT$) terms exactly annihilate any state in the scar subspaces spanned by $\ket{n^\eta}$ and $\ket{n^\eta}'$.

The decomposition for the zeta states is
\begin{gather}
\label{eq:HubDecompZeta}
H^{H} =  H_0^{H\zeta} + \sum_{\braket{kj}} T_{kj}  + \frac{U}{2} \sum\limits_i (n_i-1)^2\ ,
\end{gather}
where
\begin{gather}
\label{eq:zetaH0}
H_0^{H\zeta} = n\left(\frac{U}{2} +\mu_H\right) - U \frac{N}{2}
\end{gather}
and the remaining ($OT$) terms annihilate any state in the zeta scar subspace spanned by $\ket{n^\zeta}$. All the zeta states are half-filled \cite{Pakrouski:2021jon}, therefore the action of \eqref{eq:zetaH0} within the zeta subspace is just a constant $\mu_HN$.

The scar subspace constructed using eq. \eqref{eq:genTowerWf} may have symmetry $G$ that is higher than O$(N)$. For example the $\ket{n^\eta}$ built using $\eta^\dagger$ \eqref{eq:1bandPairingOdag} are in addition spin-singlets and their full symmetry is $G=\widetilde{Sp}(N)$ \cite{paperB}. In all three cases O$(N)$ is a sub-group of the full symmetry $G$.

\subsection{BCS ground state for type-\rom{1} $\eta$ and $\eta'$}

Now on top of the Hubbard Hamiltonian we add the pairing potential $\delta H_0$ with $\CO$ from \eqref{eq:1bandPairingOdag}. For $\eta$ and $\eta'$ one can think of the resulting Hamiltonian as being a model for a Hubbard material being brought in close proximity of a superconductor. In all the three cases this additional term respects the full symmetry of the respective scar family and becomes part of $H_0$. For example the exact full Hamiltonian that governs eta scar subspace is then
\begin{gather}
\label{eq:fullEtaH0}
H^{\eta}_0 = H^{H\eta}_0 +\delta H^\eta_0
\end{gather}
Alternatively, one could start from $H^{\eta}_0$ (which is identical to the mean-field Hamiltonian) and consider Hubbard and any further $OT$ terms as a perturbation that, independent of its strength, leaves the mean-field solution intact because it is a scar state.

Consider the two cases 1) real hopping in \eqref{eq:HimHub}, $\CO_j = {\eta '}_j$, the tower of states $\ket{\phi_n}$ is then known as momentum $k=\pi$ eta-pairing states \cite{etaPairingYang89} defined on any bi-partite lattice 2) imaginary hopping in \eqref{eq:HimHub}, $\CO_j = {\eta}_j$, states $\ket{\phi_n}$ are then $k=0$ eta-pairing states \cite{etaPairingYang89,pakrouski2020GroupInvariantScars} defined on arbitrary  lattice). Define $s_j = e^{i k j}$.
$H^{\eta}_0$ is of the general form
of eq. \eqref{eq:H0generalSU2} with $\mu=\frac{U}{2}+\mu_H$ and $C=(\frac{U}{2}+\mu_H) N$.

According to the general argument for a large enough $\gamma$ the ground state of the Hamiltonian $H^{H}+\delta H_0$ is a many-body scar state

\begin{align}
\label{eq:etaPspgs}
|z^\eta_0\rangle =  N_z \prod_j \left( 1 + \frac{e^{i\theta}}{\frac{\mu}{\gamma} + \sqrt{\frac{\mu^2}{\gamma^2} + 1} } s_j  c^\dagger_{j\uparrow} c^\dagger_{j\downarrow} \right)\ket{0}.
\end{align}
which is also the BCS wavefunction written in real space.
For $\mu_H=-\frac{U}{2}$ and $\theta=0$ it maximizes over the Hilbert space the value of the one point function $\braket{s_j  c^\dagger_{j\uparrow} c^\dagger_{j\downarrow}}= \frac{e^{-i\theta}}{2} \frac{1}{\sqrt{1+\left(\frac{U/2+\mu_H}{\gamma}\right)^2}}$ and its two-point function $\braket{\frac{s_j}{s_r} c^\dagger_{j\uparrow} c^\dagger_{j\downarrow}  c_{r\uparrow} c_{r\downarrow}}=0.25$ has the maximum value that can be achieved by a product state. For arbitrary $\mu_H$ both values are independent of the system size $N$ and site indexes $j$ and $r$ which means the state \eqref{eq:etaPspgs} has off-diagonal long-range order \cite{etaPairingYang89}. The state $|z^\eta_0\rangle$ \eqref{eq:etaPspgs} and its excitations are a linear combination of the states in the original tower.

We thus establish that the BCS wavefunction $|z^\eta_0\rangle$ is a linear combination of the eta-pairing states!

\subsection{Solving $H_0$ using Bogoliubov transformation \label{sec:ApEtaBogoliubovDeriv}}

Within the scar subspace the Hamiltonian is given by a $(NK+1)\times (NK+1)$ tri-diagonal matrix with diagonal matrix elements determined by $\mu$ and the off-diagonal ones given by $H_{n,n+1} = \gamma e^{i\theta}\sqrt{(NK-n+1)n} $. 

The $H_0$ Hamiltonian governing the scar subspace can alternatively be solved using a canonical transformation as is customary for the BCS mean field Hamiltonians.

For the spin-full type-\rom{1} excitations it is convenient to use the spinor notation that (similar to the Nambu spinor \cite{originalNambu1960}) combines particles and holes 
\begin{align}
\label{eq:genType1Spinor}
\begin{pmatrix}
c^{1\dagger}_{j\uparrow} \dots c^{K\dagger}_{j\uparrow} c^1_{j\downarrow} \dots c^K_{j\downarrow}
\end{pmatrix}
\end{align}
whereas for type-\rom{2} the spinor that combines the "magnetic particles" (spin-up) with "holes" (spin-down)
\begin{align}
\label{eq:type2spinor}
\begin{pmatrix}
c^{1\dagger}_{j\uparrow} \dots c^{K\dagger}_{j\uparrow}  c^{1\dagger}_{j\downarrow} \dots c^{K\dagger}_{j\downarrow}
\end{pmatrix}
\end{align}

With respect to the symmetry groups of the $\ket{\eta}$ and $\ket{\eta}'$ states the $H_0$ part of the Hamiltonian is given by eq. \eqref{eq:fullEtaH0}.
Let's write it using the type-\rom{1} spinor \eqref{eq:genType1Spinor} and $\mu=(U/2+\mu_H)$

\begin{align}
\label{eq:H0etaInSpinorB}
H^\eta_0 = \sum_j
\begin{pmatrix}
c^\dagger_{j\uparrow} & c_{j\downarrow}
\end{pmatrix}
\begin{pmatrix}
\mu & -\gamma s_j e^{i\theta} \\
-\gamma s_j^* e^{-i\theta} & -\mu
\end{pmatrix}
\begin{pmatrix}
c_{j\uparrow} \\ 
c^\dagger_{j\downarrow}
\end{pmatrix}
+ \mu N
\end{align}

Observe that this Hamiltonian is very similar to the BCS mean-field Hamiltonian \cite{TinkhamBookIntroSC}
\begin{align}
\label{eq:BCSHam}
H_0 = \sum_k
\begin{pmatrix}
c^\dagger_{k\uparrow} & c_{-k\downarrow}
\end{pmatrix}
\begin{pmatrix}
\xi_k & -\Delta_k \\
-\Delta^*_k & -\xi_k
\end{pmatrix}
\begin{pmatrix}
c_{k\uparrow} \\
c^\dagger_{-k\downarrow}
\end{pmatrix}
+ K_0
\end{align}

We can therefore diagonalize \eqref{eq:H0etaInSpinorB} using the Bogoliubov canonical transformation.

\begin{align}
\label{eq:H0etaDiagonalized}
H_0^\eta = E \sum_j (\gamma^\dagger_{j1} \gamma_{j1} + \gamma^\dagger_{j2} \gamma_{j2}) + (\mu- E)N=\\
\sum_j \begin{pmatrix}
\gamma^\dagger_{j1} & \gamma_{j2}
\end{pmatrix}
\begin{pmatrix}
E & 0 \\
0 & -E
\end{pmatrix}
\begin{pmatrix}
\gamma_{j1} \\ 
\gamma^\dagger_{j2}
\end{pmatrix}
+ \mu N
\end{align}

Where 
\begin{align}
\label{eq:Udef}
\begin{pmatrix}
c_{j\uparrow} \\ 
c^\dagger_{j\downarrow}
\end{pmatrix} = 
\begin{pmatrix}
u^*_j & s_j v_j\\ 
-s_j^* v^*_j & u_j
\end{pmatrix}
\begin{pmatrix}
\gamma_{j1} \\ 
\gamma^\dagger_{j2}
\end{pmatrix},
\end{align}

and the definitions of eq. (15) in the main text which we reproduce here 
 \begin{align}
\small
\label{eq:uvXdef}
 \quad u=\sqrt{\frac{1+\frac{\mu}{E}}{2}}, \quad v=e^{i\theta}\sqrt{\frac{1-\frac{\mu}{E}}{2}}, \quad
E = \sqrt{\gamma^2+\mu^2}
\end{align}
\normalsize
 
 are used. The site index has been omitted as there is no site-dependence.

The creation operators of the new emergent fermions are
\begin{align}
\label{eq:EtaTransformedFermions}
\gamma^\dagger_{j1} =  u c^\dagger_{j\uparrow} - s_j^*v^* c_{j\downarrow}; \quad
\gamma^\dagger_{j2} =  s_j^*v^* c_{j\uparrow} +  u c^\dagger_{j\downarrow} .
\end{align}

The eigenstates of \eqref{eq:H0etaDiagonalized} is a tower of $N+1$ states \eqref{eq:genTowerWf}
with the raising operator \eqref{eq:gammaRaisingOpGeneral}
alternatively constructed by replacing the original fermion operators with $\gamma^\dagger_{j}$ in eq. \eqref{eq:1bandPairingOdag}, that creates one additional quasiparticle of each type
\begin{align}
\label{eq:etaGammaRaisingOPeratorBogoliubov}
\CO^{\gamma \dagger}_j = \frac{1}{2}\sum_j s_j \gamma^\dagger_{j1} \gamma^\dagger_{j2} 
\end{align}

The lowest-energy state $\ket{z^\eta_0}$ in the tower scar subspace \eqref{eq:etaPspgs}, the BCS wavefunction, has no excitations of either type and thus satisfies for any site $j$ 
\begin{align}
\label{eq:BCSgsCondition}
\CO^{\gamma}_j  \ket{z^\eta_0} = 0
\end{align}

\subsection{Additional numerical results}

\begin{figure}[htp!]
	\begin{center}
\includegraphics[width=0.49\columnwidth]{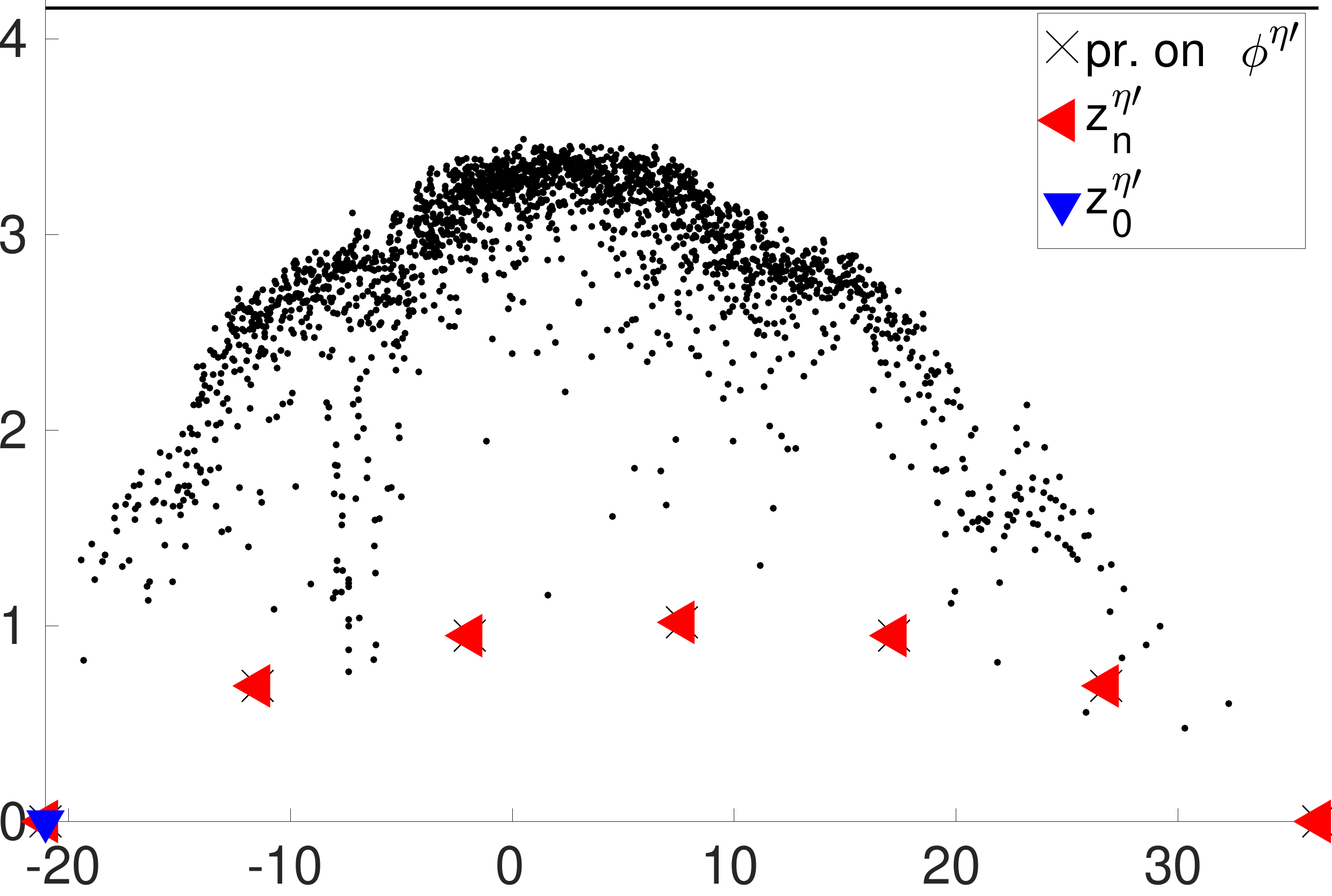}
\includegraphics[width=0.49\columnwidth]{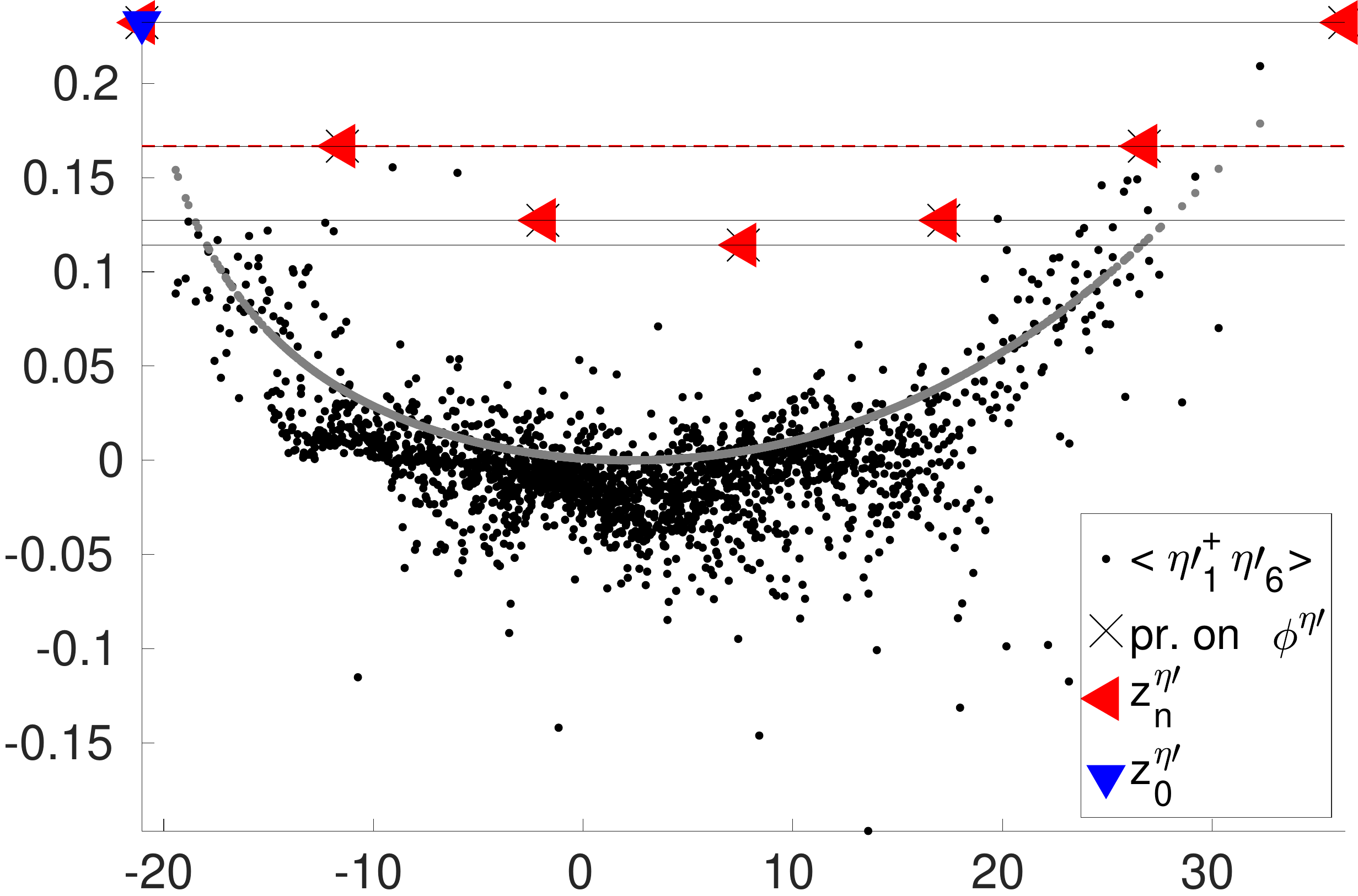}
\includegraphics[width=0.49\columnwidth]{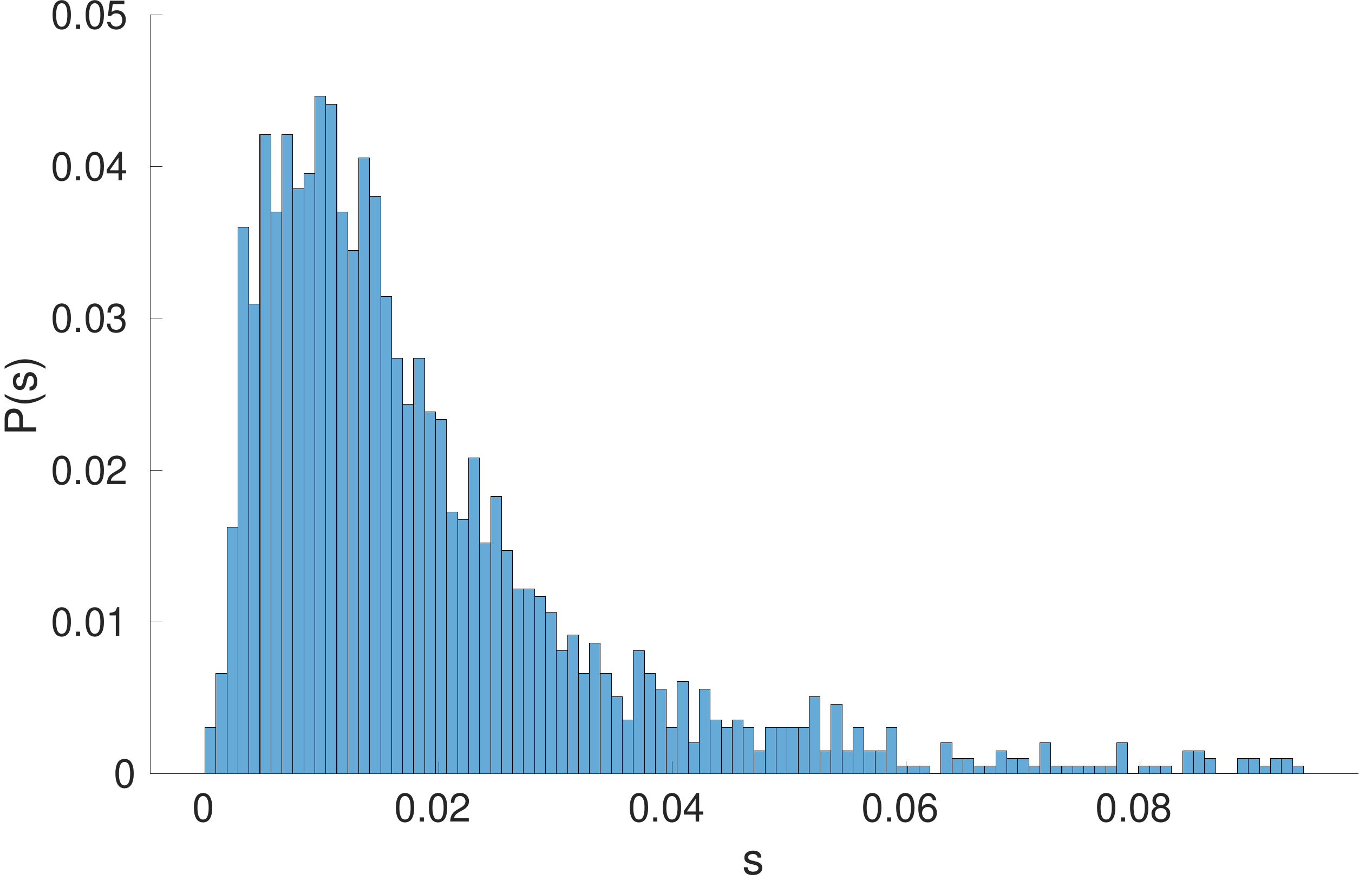}
\end{center}
\caption{\label{fig:numericsEtaPSM} Numerical results for $K=1$, 2D real-hopping Hubbard model with o.b.c., $U=5.01$, $\mu=-1.2295$, $\theta=\pi/7$, $N_x\times N_y=2\times3$ and $\gamma=4.61$. a) Entanglement entropy for a cut separating the system into 2 pieces (cut horizontally 1+2 rows). All the $\ket{z_n}$ states have unity projection on the subspace spanned by $\ket{\phi_n}$. b) ODLRO $\braket{\CO^\dagger_i\CO_j}$ with $i=1$ and $j=N$ - the two sites with largest separation in our lattice. Dashed horizontal line indicates the average over the scar subspace. c) Histogram of all the energy gaps in the spectrum, excluding the MBS states. Horizontal axis in panels a) and b) is energy.
}
\end{figure}

In Fig. \ref{fig:numericsEtaPSM} we show the data analogous to the left column of Fig. 1 in the main text with the only exception of smaller system size: $2\times3$ vs $2\times4$.

These results again agree with all the analytical predictions about the MBS subspace including the exact overlap with the analytical wavefunctions. 
While for the $2\times4$ system in the main text the generic, non-scar states are well-concentrated around the expected thermal average, here for $2\times3$, due to the smaller system size, the generic states are significantly more spread out.

\subsection{Ground state for type-\rom{2} $\zeta$}

For complex-valued hopping in \eqref{eq:HimHub} and $\CO^\dagger_j ={\zeta}_j^{\dagger}  = c^\dagger_{j\uparrow} c_{j\downarrow}$ the Hamiltonian governing the scar subspace is 
\begin{align}
\label{eq:zetaH0full}
H_0^{\zeta} = H_0^{H\zeta} + \delta H^\zeta_0.
\end{align}

It is of the general form \eqref{eq:H0generalSU2} with $\mu=0$, $C=\mu_HN$.

The lowest-energy state within the scar subspace and the large-$\gamma$ global ground state is
\begin{align}
\label{eq:zetaPspgs}
\ket{z^\zeta_0} =  2^{-\frac{N}{2}} \prod_j \left( 1 + e^{i\theta}c^\dagger_{j\uparrow} c_{j\downarrow} \right)\ket{0_{\rom{2}}}.
\end{align}
In agreement with the general expressions its one- and two-point (ODLRO) functions are $\braket{c^\dagger_{j\uparrow} c_{j\downarrow} }=0.5e^{-i\theta}$ and $\braket{c^\dagger_{j\uparrow} c_{j\downarrow} c^\dagger_{r\downarrow} c_{r\uparrow} }=0.25e^{-i2\theta}$ and are independent of the system size $N$ and of the positions where they are evaluated.

\subsection{Solving $H_0$ using Bogoliubov transformation \label{sec:AppDeriveBogoliubovForZeta}}

Using the type-\rom{2} spinor \eqref{eq:type2spinor} we write

\begin{align}
H^\zeta_0 = \sum_j
\begin{pmatrix}
c^\dagger_{j\uparrow} & c^\dagger_{j\downarrow}
\end{pmatrix}
\begin{pmatrix}
0 & -\gamma e^{i\theta} \\
-\gamma e^{-i\theta} & 0
\end{pmatrix}
\begin{pmatrix}
c_{j\uparrow} \\ 
c_{j\downarrow}
\end{pmatrix}
+ \mu N
\end{align}

The transformation is the same as in the eta case (replacing $\mu=0$ and using \eqref{eq:uvXdef})
\begin{align}
\begin{pmatrix}
c_{j\uparrow} \\ 
c_{j\downarrow}
\end{pmatrix} = 
\begin{pmatrix}
u^*_j & v_j\\ 
-v^*_j & u_j
\end{pmatrix}
\begin{pmatrix}
\gamma_{j1} \\ 
\gamma_{j2}
\end{pmatrix}
\end{align}
leading to the emergent fermion operators
\begin{align}
\label{eq:transformedZetaFermions}
\gamma^\dagger_{j1} =u c^\dagger_{j\uparrow} - v^* c^\dagger_{j\downarrow}; \quad
\gamma_{j2} = v^* c_{j\uparrow} +  u c_{j\downarrow}.
\end{align}

The Hamiltonian after the canonical transformation is
\begin{align}
\label{eq:H0zetaDiagonalized}
H^\zeta_0 = 
\begin{pmatrix}
\gamma^\dagger_{j1} & \gamma^\dagger_{j2}
\end{pmatrix}
\begin{pmatrix}
E_1 & 0 \\
0 & E_2
\end{pmatrix}
\begin{pmatrix}
\gamma_{j1} \\ 
\gamma_{j2}
\end{pmatrix}= E \sum_j \gamma^\dagger_{j1} \gamma_{j1} - \gamma^\dagger_{j2} \gamma_{j2}
\end{align}
and is the total "magnetization" wrt to the flavour of the transformed fermions.

The eigenstates of \eqref{eq:H0zetaDiagonalized} is a tower of states \eqref{eq:genTowerWf} where the raising operator
\begin{align}
\label{eq:OgammaWS19}
\CO^{\gamma \dagger}_j = \gamma^\dagger_{j1} \gamma_{j2}.
\end{align}
flips the "spin"/flavour of the transformed fermions analogous to ${\zeta}_j^{\dagger}$ \eqref{eq:1bandPairingOdag}.

\subsection{Benefits of many-body scars}

\begin{figure}[htp!]
	\begin{center}
								\includegraphics[width=0.49\columnwidth]{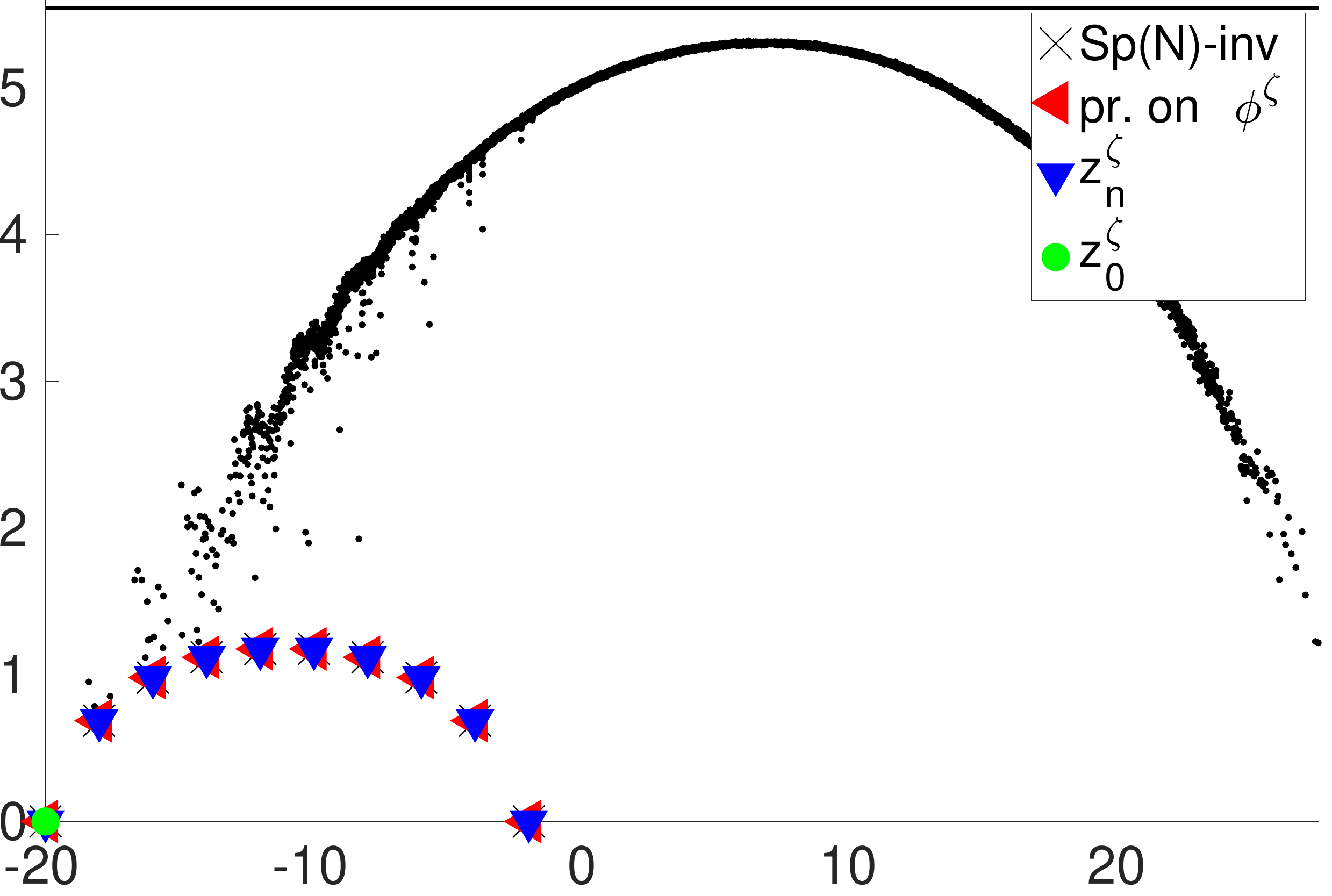}
				\includegraphics[width=0.49\columnwidth]{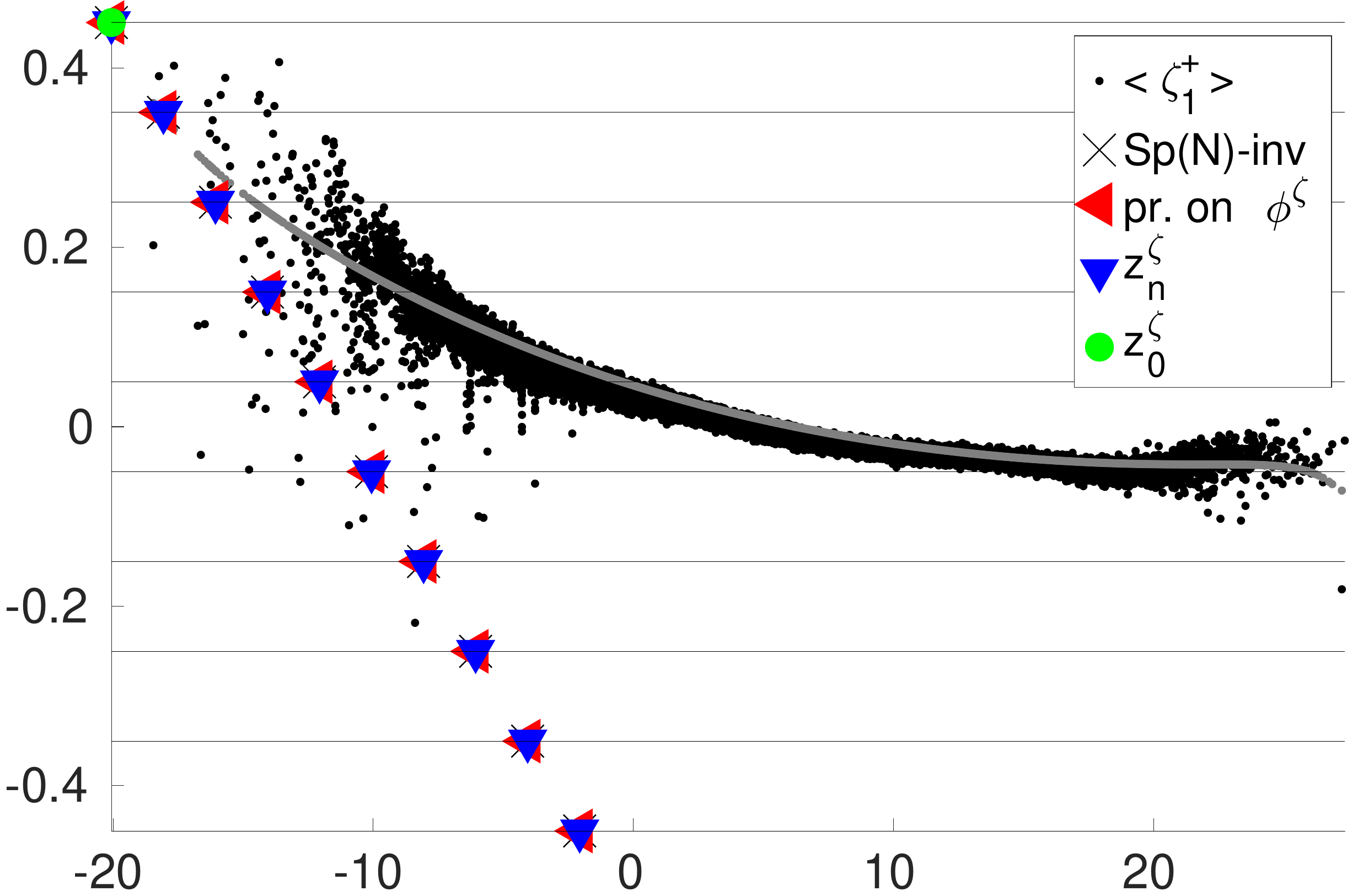}\\
								\includegraphics[width=0.49\columnwidth]{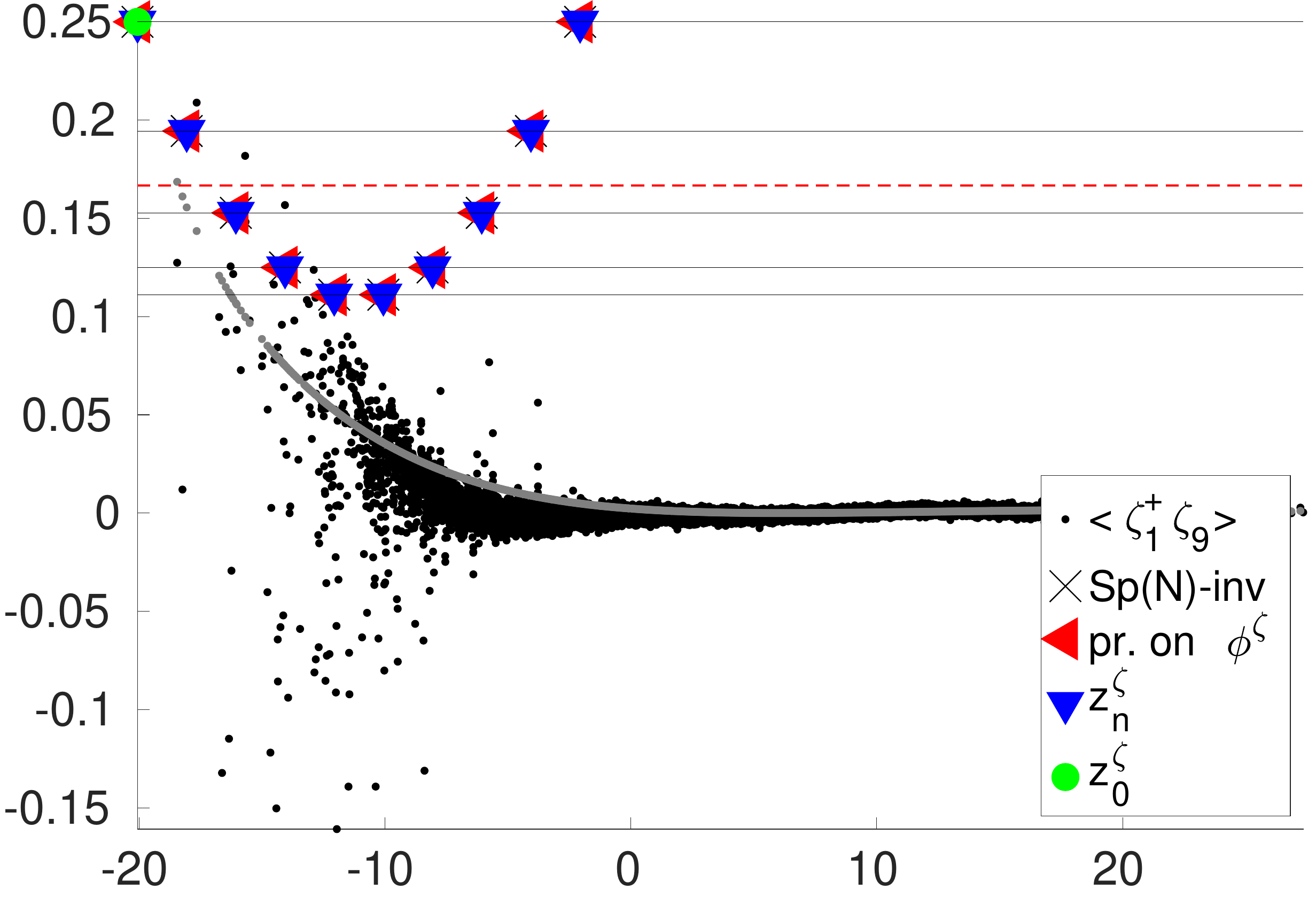}
												\includegraphics[width=0.49\columnwidth]{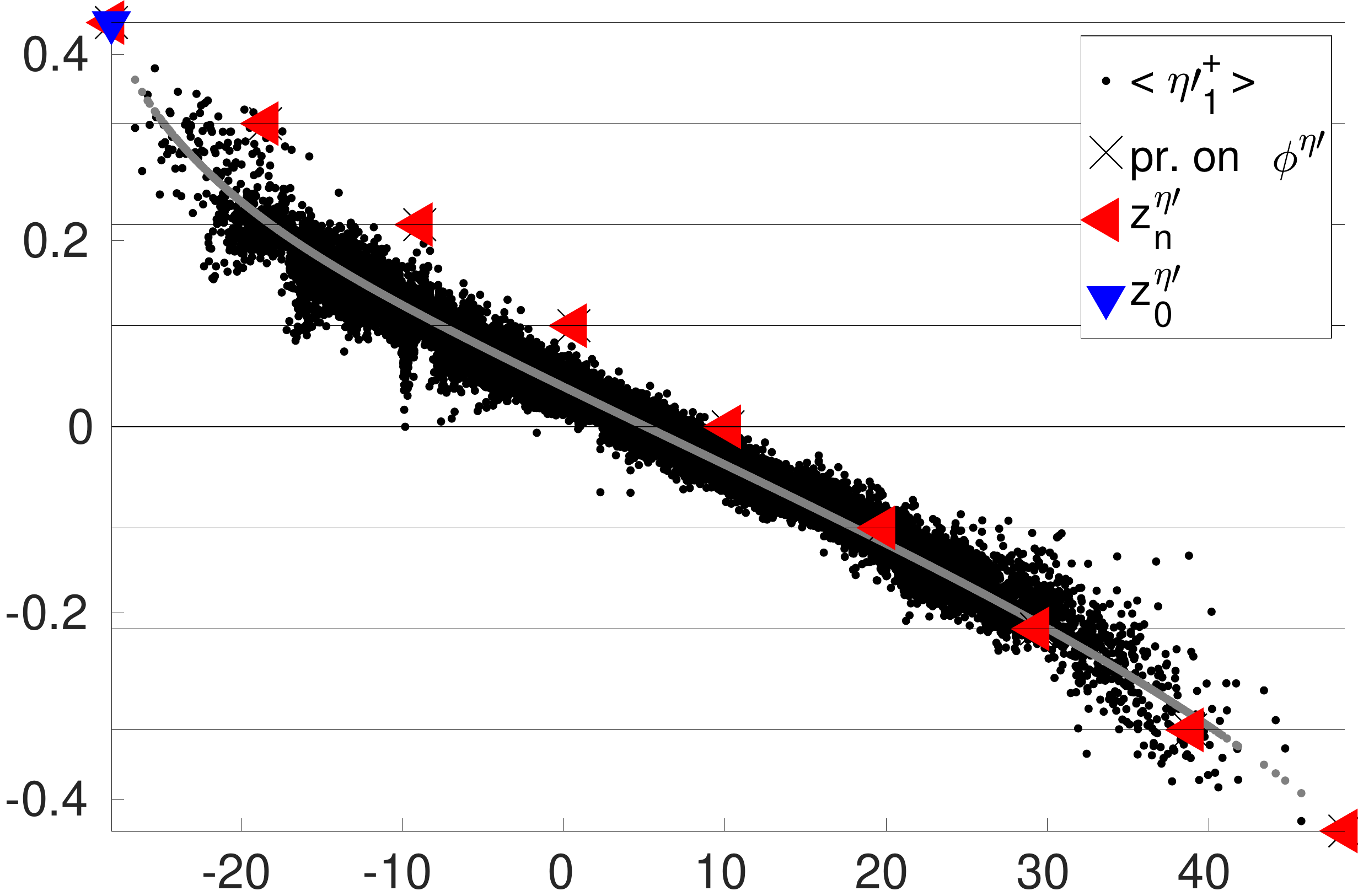}
	\end{center}
\caption{\label{fig:numericsEtaPandZeta} Numerical results for single-flavour ($K=1$) 2D real-hopping Hubbard model with o.b.c., $U=5.01$, $\theta=\pi/7$, $\mu=-1.2294$. In a),b)c) $\CO^\dagger=\zeta^\dagger$, $N_x\times N_y=3\times3$ and $\gamma=1$. In d) $\CO^\dagger=\eta^{\dagger\prime}$, $N_x\times N_y=2\times4$ and $\gamma=4.61$. 
  a) entanglement entropy b),d) one-point function $\braket{\CO^\dagger} _j$ on site (1,1). Black horizontal lines indicate the analytical prediction; c) two-point function (ODLRO) $\braket{\CO^\dagger _i\CO _j}$ where site $i$ is (1,1) and site $j$ is $(N_x,N_y)$. Dashed horizontal line indicates the average over the scar subspace. Black horizontal lines indicate analytical values. In d) the calculation is performed at fixed (even) total fermion number parity and in a),b),c) at fixed (half-filling) total particle number. Horizontal axis in all the plots is energy.
}
\end{figure}

Because $|z_0\rangle $ and its $\CO^{\gamma\dagger}$ excitations are scar states our results hold for any deformed Hamiltonian 
\begin{align}
\label{eq:Hdeformed}
\tilde{H} = H^{H}+\delta H_0 + \sum_l O_l T_l,
\end{align}
where $O_l$ is an arbitrary operator and $T_l$ is a generator of the symmetry group of one of the three scar families. Ref. \cite{Pakrouski:2021jon} provides an exhaustive list of such generators. They include arbitrary-range hopping terms, spin-orbit coupling (for the eta states $\CO_j = {\eta}_j$), magnetic field (for eta and eta$^\prime$) and many others. Ref. \cite{Pakrouski:2021jon} also shows that many commonly used interactions such as Hubbard and Heisenberg also decompose as $H_0+OT$. Thus such interactions might at most shift the values of $\mu$ and $\gamma$ but usually only contribute $OT$ terms that are irrelevant within the scar subspace.

We confirm our analytical conclusions by diagonalizing exactly numerically small 2D systems with the results for the eta$^\prime$ and zeta states shown in Fig. \ref{fig:numericsEtaPandZeta} and in panels a),c),e) in Fig. 1 in the main text. To illustrate the stability of our solution \eqref{eq:Hdeformed} we choose the $OT$ term given in eq. (23) in the main text.
We choose $t=e^{i\sqrt{2}\pi}$ for the zeta states and $t=1$ for the eta$^\prime$ states. These additional $OT$ terms are intended to simulate some quite drastic and strong perturbation that can be added to the Hubbard model without disturbing the BCS scar ground state and its excitations. We also choose it to be quite chaotic such that it breaks most symmetries of the full Hamiltonian.

Within the total fermion number parity (or half-filling for zeta states) sector the full Hamiltonian \eqref{eq:Hdeformed}, indeed has no other symmetries remaining. The bulk of its spectrum (excluding MBS) is fully ergodic with the level statistics parameter (average ratio of consecutive energy gaps \cite{Atas_2013}) $\braket{r}=0.5465$ (for $\eta^\prime$) which is rather close to 0.5359 expected \cite{Atas_2013} in a generalized orthogonal ensemble (GOE, real, symmetric, random matrix). The ergodicity of the spectrum bulk is also confirmed by the energy gap distribution that follows the one expected for GOE and exhibits absence of near-zero gaps as a consequence of level repulsion.

The ground state $\ket{z_0}$ \eqref{eq:zetaPspgs} (green circle in Fig. \ref{fig:numericsEtaPandZeta}) and the excitations $(\CO^{\gamma\dagger})^n\ket{z_0}$ above it (blue triangles) are identified by an exact identity overlap with the analytical wavefunctions. In all the cases that we have studied numerically the ground state has the largest (over the full Hilbert space) expectation value for both 1- $\braket{\CO^{\dagger}_j}$ and 2-point $\braket{\CO^{\dagger}_i \CO_j}$ function also at finite $\mu$ where we do not have an analytical argument guaranteeing this. Furthermore MBS spanned by \eqref{eq:genTowerWf} have higher $\braket{\CO^{\dagger}_i \CO_j}$ (ODLRO) as a subspace. Indeed, the average absolute value of ODLRO over the eta$^\prime$ states (Fig. 1c in the main text) is 14.2 times larger than the average value over the remaining states. Similarly for the zeta states this factor is 64! Further, as can be seen in Fig. \ref{fig:numericsEtaPandZeta} c) and Fig. 1c in the main text almost none of the generic states has ODLRO exceeding the scar subspace average (dashed line).This shows that although adding the pairing potential $\delta H_0$ does induce the corresponding fluctuations in all the generic states, ODLRO in the scar subspace remains significantly larger.

Red triangles in Fig. \ref{fig:numericsEtaPandZeta}a),b),c) indicate the states that fully lie within the subspace spanned by \eqref{eq:genTowerWf} with the original raising operator \eqref{eq:1bandPairingOdag}. Blue triangles indicate the states created by the transformed raising operator $\CO^{\gamma \dagger}_j$ (eq. \eqref{eq:OgammaWS19}) above the ground state $|z^\zeta_0\rangle$. The fact that the two groups of states coincide confirms that this ground state and the "mean-field" excitations above it form a new basis in the original scar subspace \eqref{eq:genTowerWf}. For the analogous case of eta$^\prime$ states in the main text this confirms that the BCS wavefunction and its excitations are linear combinations of the eta-pairing states.

\subsection{What $\gamma$ is large enough?}

At a relatively high $U=8$ and low $\gamma$ of about 0.69 the spectrum separates into the Mott "lobes" corresponding to fixed Hubbard-U energy that are broadened by the hopping and pairing potential. If Hubbard is repulsive the lobe containing the eta and eta$^\prime$ subspaces is highest in energy. Therefore to make the BCS wavefunction the ground state the value of $\gamma$ roughly needs to exceed the bandwidth of all other Hamiltonian terms. In our numerical experiments with unity hopping amplitude the required $\gamma_c$ is about the same as the Hubbard $U$. For attractive Hubbard the eta/eta$^\prime$ subspace is in the lowest energy "lobe" and to make BCS state the ground state $\gamma_c$ only needs to exceed the bandwidth of that lobe. Numerically we observe that the required $\gamma_c$ is about $U/10$.

\section{Example 2: Inter-orbital magnetism in a two-orbital generalized Hubbard model \label{sec:ExIBZeta}}

Consider an example of the type-\rom{2} Hilbert space with $K=2$ orbitals or fermion flavours per site: $A$ and $B$ and the inter-orbital magnetic excitation creation operator 
\begin{align}
\label{eq:ibZetaOdag}
\CO^{\dagger}_j= c^{A\dagger}_{j\uparrow}c^B_{j\downarrow} + c^{B\dagger}_{j\uparrow} c^A_{j\downarrow}
\end{align}
that couples the spin and orbital degrees of freedom.

We consider the generalized Hubbard Hamiltonian given by eq. (25) in the main text
in external magnetic field $B_z$ and with spin-orbit coupling (main text eq. (26)).

The hopping amplitudes $t_{kj}$ in the second term are imaginary and are thus generators of O$(N)$ \cite{Pakrouski:2021jon}). The generalized 2-orbital Hubbard interaction \cite{paperB} $\sum_j H_j^{\zeta Hub}$ with
\begin{align}
\label{eq:zetaHub}
H_j^{\zeta Hub} = \frac{1}{2} (n_j^2 - n_j) = \frac{1}{2} n_j (n_j - 1),
\end{align}
acts as a constant on any uniformly half-filled states including the scar subspace spanned by $\ket{\phi_n}$ \eqref{eq:genTowerWf} with $\CO$ from \eqref{eq:ibZetaOdag}. In the single-flavour (K=1) case this interaction is identical to the standard Hubbard interaction. In the present two-flavour case (K=2) it can also be written \cite{paperB} as
\begin{align}\label{JUK}
H_j^{\zeta Hub} =  \sum_{ \sigma} n^A_{j\sigma} n^B_{j\sigma}+ \sum_{p} n^p_{j\uparrow}n^p_{j \downarrow}+\sum_{p, q} n^p_{j\uparrow}\sigma^1_{pq}n^q_{j\downarrow}~,
\end{align}
which is a special case of the on-site Coulomb interaction for $\rm{Sr_2RuO_4}$ considered in \cite{typicalUSCHamZinkl2021}.

The particle number operator is

\begin{align}
\label{eq:2BdensityOp}
n_j = n_{j\uparrow} + n_{j\downarrow} = \sum^B_{p=A} c^{p\dagger}_{j,\uparrow}c^{p}_{j,\uparrow} + \sum^B_{p=A} c^{p\dagger}_{j,\downarrow}c^{p}_{j,\downarrow}
\end{align}

\begin{figure}[htp!]
	\begin{center}
				\includegraphics[width=0.49\columnwidth]{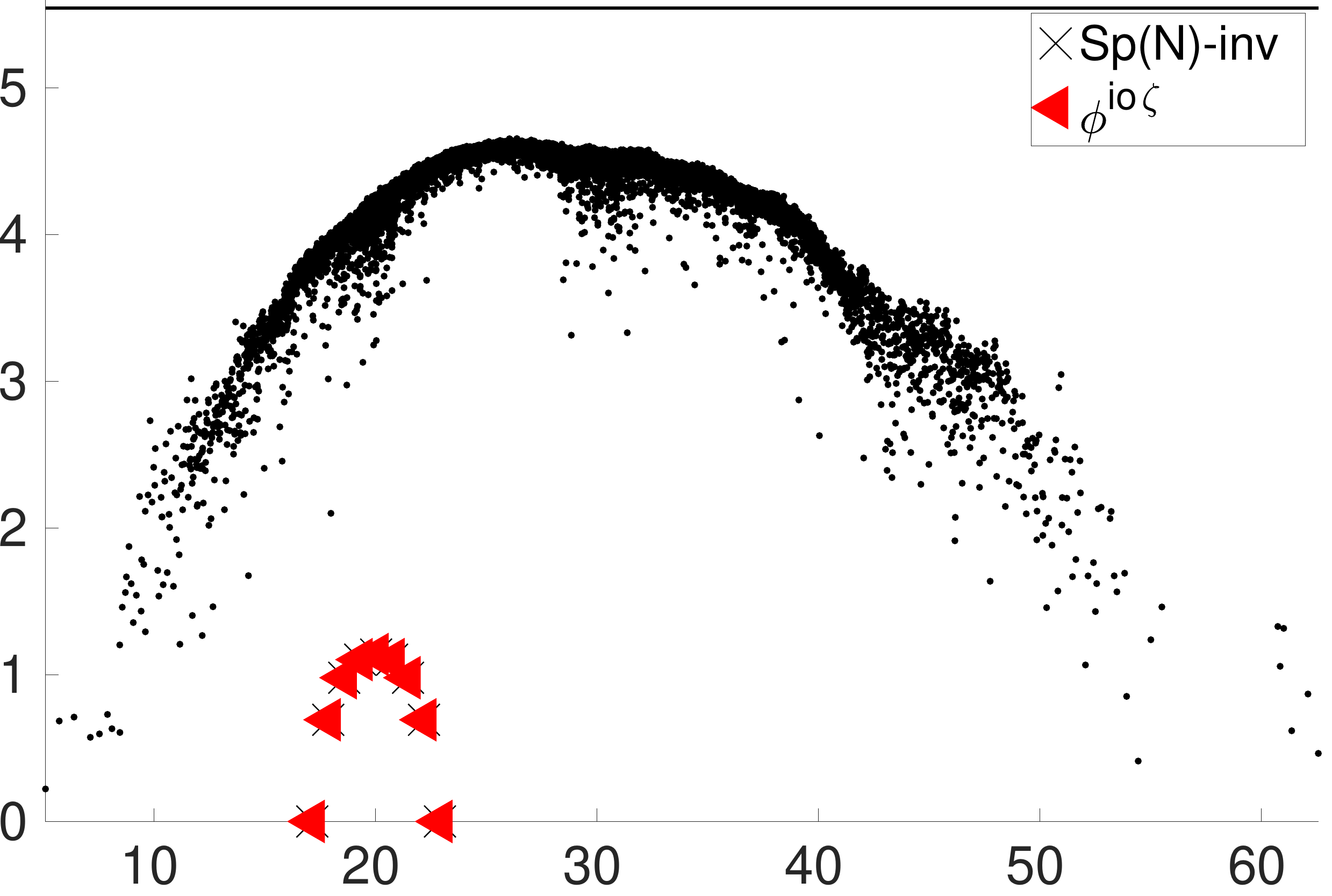}		
				\includegraphics[width=0.49\columnwidth]{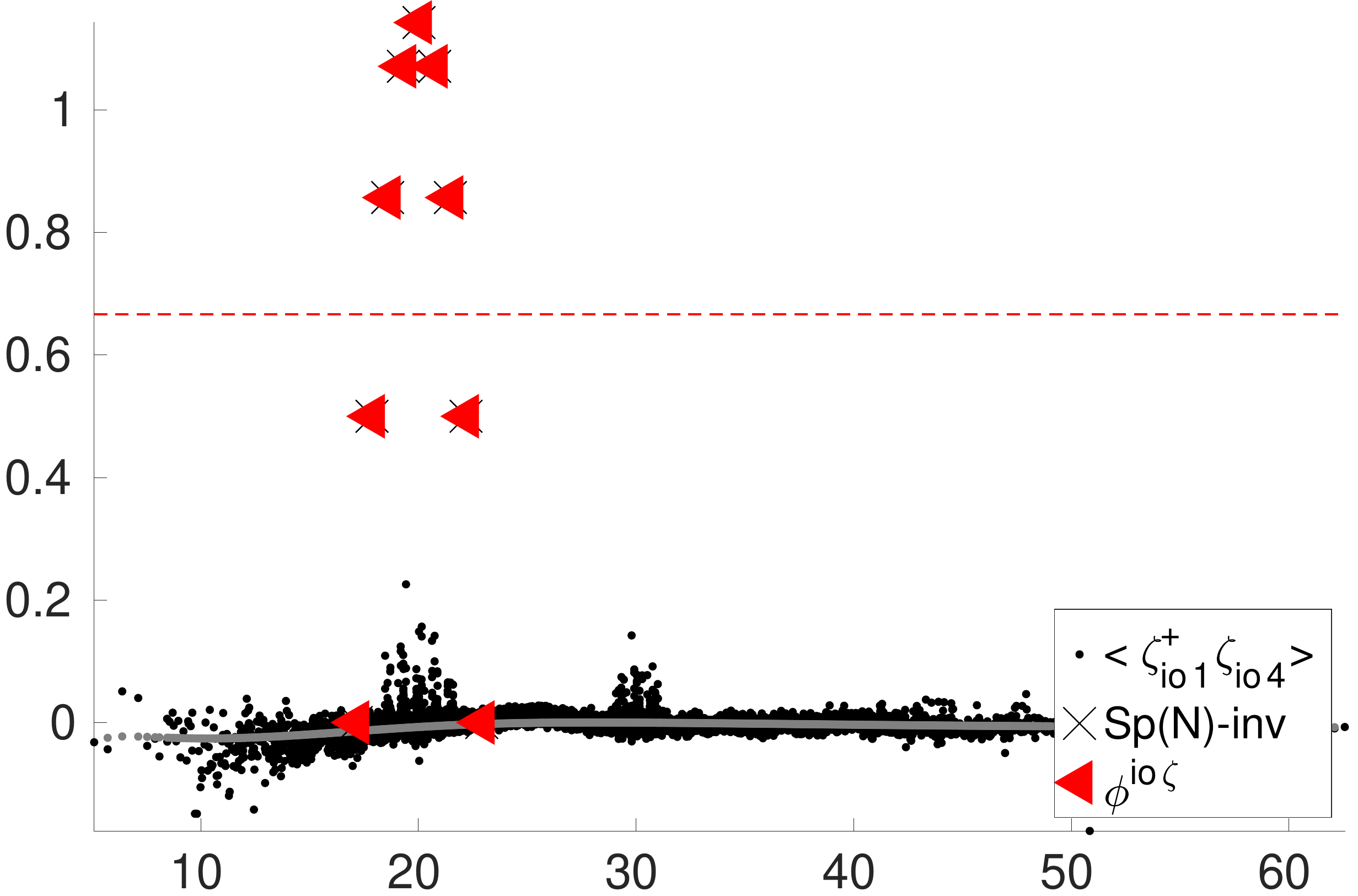}
	\end{center}
\caption{\label{fig:ibZetaWithoutGamma} Numerical results for inter-orbital zeta $\CO$ \eqref{eq:ibZetaOdag} and the generalized Hubbard Hamiltonian \eqref{eq:HibZdeformed} without the $\delta H_0$ term. $K=2$, one dimension, $N=4$, $t=i$, $B^p_z=\{0.3147; 0.4058\}$, $\theta=\pi/7$, $U=5.01$, $\gamma=2.5$, fixed total particle number (half-filling).a) Entanglement entropy b) 2-point function}
\end{figure}

Fig. \ref{fig:ibZetaWithoutGamma} shows the inter-band zeta states in the basis \eqref{eq:genTowerWf} without the $\delta H_0$ term added to the Hamiltonian.

Over the MBS subspace the $H_0$ part of the Hamiltonian consists of the first and the last terms of the Hamiltonian (eq. (25) in the main text) 
 \begin{align}
\label{eq:ibZetaH0}
H_0^{io\zeta} = \sum_{i=1}^N \sum_{p=A}^B B^p_z (n^p_{i\uparrow}- n^p_{i\downarrow})  - \gamma \sum_j ( e^{i\theta} \CO^{\dagger}_j + e^{-i\theta} \CO_j),
\end{align}
where we omitted the constant coming from the Hubbard interaction.

Therefore the general solution applies with $\mu=0.5(B_z^A+B_z^B)$ and the lowest energy scar state
\begin{align}
\label{eq:gsIBZeta}
|z^{io\zeta}_0\rangle =  N_z \prod_j \left( 1 + \frac{v}{u}\CO_j^\dagger + \frac{ (\frac{v}{u}\CO_j^\dagger)^2 }{2}  \right)\ket{0_{\rom{2}}}.
\end{align}

\begin{figure}
	\begin{center}
				\includegraphics[width=0.49\columnwidth]{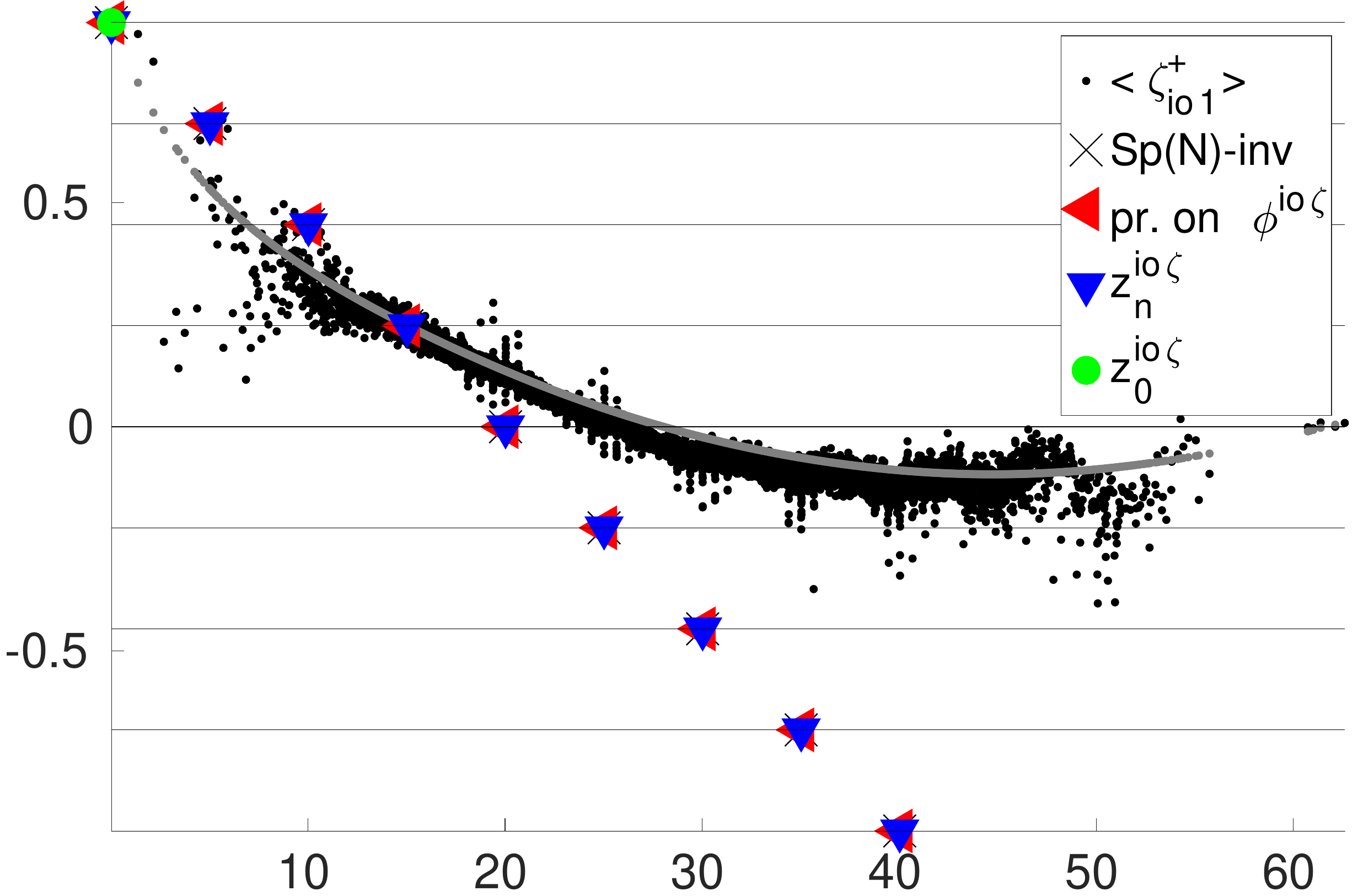}		
				\includegraphics[width=0.49\columnwidth]{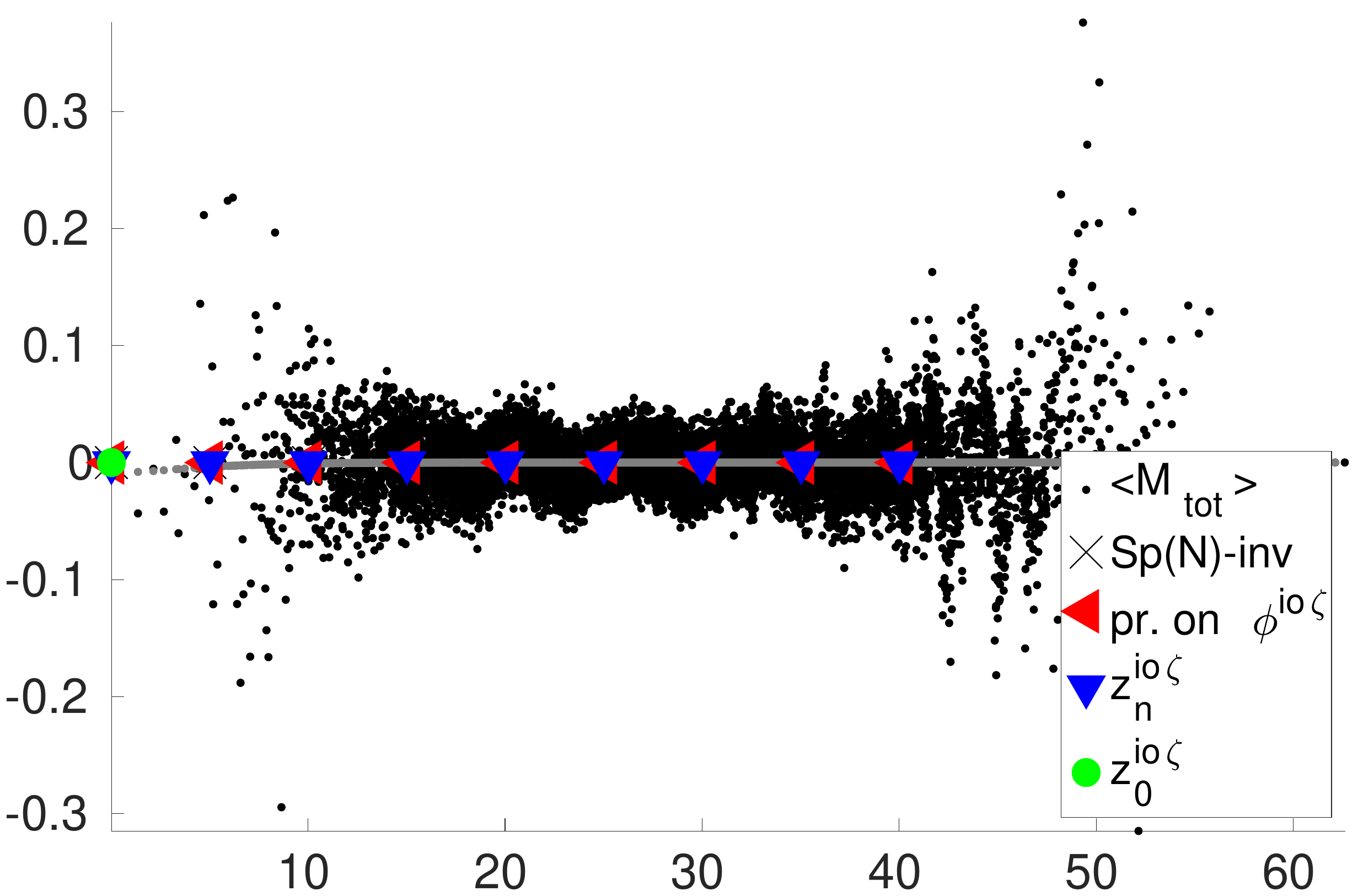}
	\end{center}
\caption{\label{fig:ibZeta} Additional numerical results for inter-orbital zeta $\CO$ \eqref{eq:ibZetaOdag} and the generalized Hubbard Hamiltonian \eqref{eq:HibZdeformed} for the same parameters as in Fig. 1
 in the main text. a) One-point function $\braket{\CO^\dagger_j}$ at site $j=1$. Horizontal black lines indicate the analytical values.
 b) Total magnetization. Horizontal axis in both plots is energy. }
\end{figure}

For the numerical calculations presented in Fig. \ref{fig:ibZeta} and main text Fig. 1
 we again consider the Hamiltonian where a perturbation (main text eq. (23))
 with $t=i$ (that has no effect on the scar subspace) is added to the Hamiltonian

\begin{align}
\label{eq:HibZdeformed}
H^{io\zeta} = H^{B}+\delta H_0 + \sum_l O_l T_l,
\end{align}

Numerical results presented in Fig. \ref{fig:ibZeta} confirm that the inter-orbital tower of O$(N)$-symmetric states are MBS. Similar to the first example we observe that the ODLRO in the scar subspace is significantly larger (factor of 7 on average) than in generic states.

In the type-\rom{2} spinor \eqref{eq:type2spinor} basis the $H_0$ part of the Hamiltonian reads
\begin{align}
\small
\begin{pmatrix}
c^{A\dagger}_{j\uparrow} & c^{B\dagger}_{j\uparrow} & c^{A\dagger}_{j\downarrow} & c^{B\dagger}_{j\downarrow}
\end{pmatrix}
\begin{pmatrix}
B_z 				& 	0				& 	0				&	-\gamma e^{i\theta} \\
0				&	B_z 				&	-\gamma e^{i\theta}	&  	0	 \\
0				&	-\gamma e^{-i\theta}	&	-B_z 				&  	0	 \\
-\gamma e^{-i\theta}	&	0				&	0 				&  	-B_z	 \\
\end{pmatrix}
\begin{pmatrix}
c^{A}_{j\uparrow} \\
c^{B}_{j\uparrow} \\
c^{A}_{j\downarrow} \\
c^{B}_{j\downarrow}
\end{pmatrix}
\normalsize
\end{align}

it is diagonalized by the following unitary transformation
\begin{align}
U = 
\begin{pmatrix}
u^*	&	0	&	v 	& 	0	& 	\\
0	&  	u^*	 &	0	&	v 	&	\\
0 	&  	-v^*	 &	0	&	u	&	\\
-v^* 	&  	0	 &	u	&	0	&	\\
\end{pmatrix}
\end{align}
where $u$ and $v$ are given by \eqref{eq:uvXdef}.

Therefore the new fermionic operators (with "orbital" index $C$,$D$ and "spin" index 1,2) are

\begin{align}
\label{eq:ibZetaGammaFermions}
\gamma^{C\dagger}_{j1} = u^* c^{A\dagger}_{j\uparrow} - v^*c^{B\dagger}_{j\downarrow}\\
\gamma^{D\dagger}_{j1} = u^* c^{B\dagger}_{j\uparrow} - v^* c^{A\dagger}_{j\downarrow} \\
\gamma^{C}_{j2} = v^* c^A_{j\uparrow} + u^* c^B_{j\downarrow} \\
\gamma^{D}_{j2} = v^* c^B_{j\uparrow} + u^*c^A_{j\downarrow}
\end{align}

in terms of which the Hamiltonian $H^{io\zeta}$ within the scar subspace is simply the total magnetization with respect to the "spin" of the transformed fermions
\begin{align}
H_0^{io\zeta} = E\sum_{p=C}^D( n^p_1 - n^p_2 ) 
\end{align}

The raising operator \eqref{eq:gammaRaisingOpGeneral} in the transformed tower that creates excitations above the ground state \eqref{eq:gsIBZeta} coincides with the original raising operator \eqref{eq:ibZetaOdag} written in terms of the Bogoliubov-transformed fermions
\begin{align}
\label{eq:ibZetaTransformedRaisingOp}
\CO^{\gamma\dagger} = \sum_j \gamma^{C\dagger}_{j1}\gamma^{D}_{j2} + \gamma^{D\dagger}_{j1}\gamma^{C}_{j2}
\end{align}

\bibliography{scar}

\end{document}